\documentclass[a4paper,fleqn,usenatbib,useAMS]{mnras}

\usepackage{graphicx}
\usepackage{amsmath}
\usepackage{amssymb}
\usepackage{bm}
\usepackage{pdflscape}
\usepackage{float}
\usepackage[T1]{fontenc}
\usepackage{ae,aecompl}
\usepackage{color}
\usepackage{caption}
\usepackage[shortlabels]{enumitem}

\pdfoutput=1

\newcommand{\hatP}{{\hat P}}
\newcommand{\hatgamma}{{\hat \gamma}}
\newcommand{\tildeE}{{\tilde E}}
\newcommand{\real}{{\:\rm{Re}}}
\label{key}

\title[Dynamical Tides in Highly Eccentric Binaries]{Dynamical Tides in Highly Eccentric Binaries: Chaos, Dissipation and Quasi-Steady State}

\author[M. Vick and D. Lai]{Michelle Vick$^{1}$, Dong Lai$^{1}$\\$^{1}$Cornell Center for Astrophysics and Planetary Science, Department of Astronomy, Cornell University, Ithaca, NY 14853, USA}

\pubyear{2018}

\begin{document}


\label{firstpage}
\pagerange{\pageref{firstpage}--\pageref{lastpage}}
\maketitle

\begin{abstract}
Highly eccentric binary systems appear in many astrophysical contexts,
ranging from tidal capture in dense star clusters, precursors of
stellar disruption by massive black holes, to high-eccentricity
migration of giant planets.  In a highly eccentric binary, the tidal
potential of one body can excite oscillatory modes in the other during
a pericentre passage, resulting in energy exchange between the modes
and the binary orbit.  These modes exhibit one of three behaviours over
multiple passages: low-amplitude oscillations, large amplitude
oscillations corresponding to a resonance between the orbital
frequency and the mode frequency, and chaotic growth, with the mode energy reaching a level comparable to the orbital binding energy. We study these
phenomena with an iterative map that includes mode dissipation, fully exploring how the mode
evolution depends on the orbital and mode properties of the system. 
The dissipation of mode energy drives the system toward a quasi-steady state, with gradual orbital decay punctuated by resonances. We quantify the quasi-steady state and the long-term evolution of the system. A newly captured star around a black hole can
experience significant orbital decay and heating due to the chaotic growth of the mode
amplitude and dissipation. A giant planet pushed into a high-eccentricity
orbit may experience a similar effect and become a hot or warm Jupiter.
\end{abstract}


\section{Introduction}
Highly eccentric binaries appear in a variety of astrophysical
contexts. In dense stellar clusters,
two stars can be captured into a bound orbit with each
other if a close encounter transfers enough energy into stellar
oscillations \citep{Fabian75, Press77, Lee86}.
Such tidally captured binaries are highly eccentric and often involve
compact objects (black holes and neutron stars). Massive black holes
in nuclear star clusters may tidally capture normal stars, and could
build up significant masses through successive captures \citep{Stone17}. 
Indeed, stars on highly eccentric orbits around massive black holes
could be precursors of tidal disruption events \citep{Rees88},
dozens of which have already been observed, e.g. \citet{Stone16,Guillochon}.
Heating from tidal dissipation may affect the structure of stars moving
toward disruption and potentially alter the observational signal of these
events \citep{Vick17,MacLeod14}.
In exoplanetary systems, hot and warm Jupiters may be formed through high-eccentricity
migration, in which a giant planet is pushed into a highly eccentric orbit
by the gravitational perturbation from a distant companion (a star or planet); 
at periastron, tidal dissipation in the planet reduces the orbital energy, leading to
inward migration and circularization of the planet's orbit
\citep{Wu03, Fabrycky07, Nagasawa08, Petrovich15, Anderson16, Munoz16}. 
Finally, the {\it Kepler} spacecraft has revealed a class of 
high-eccentricity stellar binaries with short orbital periods, whose
light curves are shaped by tidal distortion and reflection at periastron
\citep{Thompson12, Beck14, Kirk16}; 
many of these ``heartbeat stars'' also exhibit signatures of
tidally induced stellar oscillations \citep{Welsh11, Fuller12a, Burkart12, Fuller17}. 

In a highly eccentric binary, dynamical tidal interaction occurs mainly near
pericentre and manifests as repeated tidal excitations of
stellar oscillation modes.  Since tidal excitation depends on the
oscillation phase, the magnitude and direction of the energy transfer
between the orbit and the modes may vary from one pericentre passage to
the next \citep{Kochanek92}.
Earlier works in the context of tidal-capture binaries have shown that for some combinations of orbital
parameters, the energy in stellar modes may behave chaotically and
grow to very large values. \citet{Mardling95a,Mardling95b} first uncovered this phenomenon in numerical integrations of forced stellar oscillations and orbital evolution and explored the conditions for chaotic behaviour via Lyapunov analysis.
In a later work, \citet{Mardling01} presented an empirical fitting formula for the location of a ``chaos boundary," beyond which tighter and more eccentric binaries exhibit chaotic orbital evolution. The possibility of chaotic growth of mode
energy was also explored analytically in \citet{IP04,IP07,IP11} in the context of giant
planets on eccentric orbits. On the other hand, it is also expected that
the long-term evolution of the binary depends on how effectively the
binary components can dissipate energy \citep{Kumar96}.  
Indeed, in the presence of dissipation, the system may reach a
quasi-steady state in which the orbit-averaged mode energy remains
constant \citep{Lai96, Lai97, Fuller12a}. Numerical results from \citet{Mardling95b} have shown that chaotically evolving systems will eventually settle into a quiescent state of orbital evolution when modes are allowed to dissipate.
The properties of this quasi-steady state that emerges from a chaotic dynamical system are unclear.

Given the important role played by dynamical tides in various
eccentric stellar/planetary binary systems, a clear understanding of
the dynamics of repeated tidal excitations of oscillation modes and
the related tidal dissipation is desirable.  In this paper, we develop
an iterative map (Section \ref{mapping}) that accurately captures the dynamics and dissipation
of the coupled ``eccentric orbit + oscillation modes'' system.  Using
this map, we aim to (i) characterize the classes of behaviours
exhibited by eccentric binaries due to dynamical tides, (ii) explore
the orbital parameters that lead to these behaviours, and (iii) study
how the inclusion of mode damping affects the evolution of the
system.
As we shall see,  the coupled ``eccentric orbit + oscillation modes" system exhibits a richer sets of behaviours (see Section \ref{NoDissipation}) than 
recognized in the previous works by \citet{Mardling95a, Mardling95b} and \citet{IP04}. In particular, the regime of chaotic mode growth (assuming a single mode) is 
determined by two dimensionless parameters (see Fig. \ref{fig:PhaseSpace} below), not one.
Resonances between the oscillation mode and orbital motion can significantly influence the chaotic boundary for mode growth. In the presence of mode dissipation, we show that even a chaotic system eventually reaches a quasi-steady state (Section \ref{Dissipation});  we quantify the properties of the quasi-steady state and the long-term evolution of the system. In Section \ref{MultipleModes}, we generalize our analysis to multi-mode systems. 

The results of this study are applicable to a variety of
systems mentioned at the beginning of this section. Some of these applications are briefly discussed in Section \ref{Summary}. Of particular interest is the possibility that, in the high-eccentricity migration scenario of hot Jupiter formation, chaotic mode growth, combined with non-linear damping, may lead to efficient formation of warm Jupiters and hot Jupiters. 
	
\section{Iterative Map for Mode Amplitudes} \label{mapping}
	 
Consider a binary system consisting of a primary body $M$ (a star or
planet) on an eccentric orbit with a companion $M'$ (treated as a
point mass). Near pericentre, the tidal force from $M'$ excites
oscillations in $M$. When the oscillation amplitudes are sufficiently
small, we can follow the evolution of the modes and the orbit using
linear hydrodynamics. For highly eccentric orbits ($1-e\ll 1$), the
orbital trajectory around the pericentre remains almost unchanged even
for large changes in the binary semi-major axis. Under these
conditions, the full hydrodynamical solution of the system can be
reduced to an iterative map (see Appendix \ref{sec:ApRealModel}). 
We present the following map for a single-mode system and will discuss later the
effects of multiple modes.
	 
We define the dimensionless mode energy and binary orbital energy 
in units of $|E_{B,0}|$ (the initial binary orbital energy), i.e., $\tildeE=E/|E_{B,0}|$.
Consider a single mode of the star with frequency $\omega$ and (linear) damping rate $\gamma$.
Let $a_{k-1}$ be the mode amplitude just before the $k$-th pericentre passage.
Immediately after the $k$-th passage, the mode amplitude becomes
\begin{equation}
a_{k-}=a_{k-1}+\Delta a,
\label{eq:a-}
\end{equation}
where $\Delta a$ (real) is the mode amplitude change in the ``first'' passage (i.e., when there is no pre-existing oscillation). 
We normalize $a_k$ such that the (dimensionless) 
mode energy just after $k$-th passage is $\tildeE_{k-}=|a_{k-}|^2$. Thus 
the energy transfer to the mode in the $k$-th passage is
\begin{align}\label{eq:DeltaE}
\Delta\tildeE_k &=|a_{k-}|^2-|a_{k-1}|^2 = |a_{k-1}+\Delta a|^2 - |a_{k-1}|^2.
\end{align}
In physical units, the energy transfer in the ``first" passage is given by $\Delta E_1 =(\Delta a)^2|E_{B,0}|$. The binary orbital energy ($\tildeE_{B,k}$) immediately after the $k$-th passage is given by
\begin{equation}
\tildeE_{B,k} = \tildeE_{B,k-1} -\Delta\tildeE_k =\tildeE_{B,0}-\sum_{j=1}^k\Delta\tildeE_j,
\end{equation}
and the corresponding orbital period is
\begin{equation}  \label{eq:Kepler}
{P_k\over P_0}=\left({\tildeE_{B,0}\over\tildeE_{B,k}}\right)^{3/2},
\end{equation}
where $\tildeE_{B,0}=-1$.
The mode amplitude just before the $(k+1)$-th passage is 
\begin{equation}
a_k=a_{k-}\,\text{e}^{-(\text{i}\omega+\gamma) P_k}=
\left(a_{k-1}+\Delta a\right)\,\text{e}^{-(\text{i}+\hat\gamma)\hatP_k},
\label{eq:ak}
\end{equation}
where we have defined the dimensionless damping rate and orbital period
\begin{equation}
\hatgamma ={\gamma\over\omega}, \quad \hatP_k=\omega P_k.
\end{equation}
Equations~(\ref{eq:a-})-(\ref{eq:ak}) complete the map from one orbit to the next, 
starting from the initial condition $a_0=0$,
$\tildeE_0=0$, $\tildeE_{B,0}=-1$. In the absence of dissipation, this map reduces to that of \citet{IP04}. 
The map depends on three parameters:
\begin{align}
\hatP_0 &\equiv \omega P_0, \label{eq:defP0hat}\\
|\Delta\hatP_1| & \equiv \omega |\Delta P_1| \simeq{3\over 2}\hatP_0(\Delta a)^2 = \frac{3}{2}\hatP_0\left(\frac{\Delta E_1}{|E_{B,0}|}\right), \label{eq:defDelPhat}\\
\hatgamma &={\gamma\over\omega}={\gamma P_0\over\hatP_0}={1\over\hatP_0}
\left({P_0\over t_{\rm damp}}\right).\label{eq:defgammahat}
\end{align}

To relate $\hatP_0$ and $|\Delta\hatP_1|$ to the physical parameters of the system, we scale the mode
frequency $\omega$ to $\Omega_{\rm{peri}} \equiv (GM_t/r_{\rm{peri}}^3)^{1/2}$ (where $M_t=M+M'$, and $r_{\rm{peri}}$ is the pericentre distance),  and find
\begin{equation}
\hatP_0 = \frac{2 \pi (\omega/\Omega_{\rm{peri}})}{(1-e)^{3/2}}. \label{eq:hatP}
\end{equation}
The parameter $|\Delta\hatP_1|$ is related to the energy transfer in the ``first'' pericentre 
passage via $|\Delta\hatP_1|/\hatP_0 = (3/2)|\Delta E_1/E_{B,0}|$.
If we scale $r_{\rm{peri}}$ by the tidal radius $r_{\rm tide}=R(M_t/M)^{1/3}$
(where $R$ is the stellar radius), i.e., $\eta=r_{\rm{peri}}/r_{\rm tide}$, we have, for $l=2$,
\begin{equation}
\Delta E_1=-{G{M'}^2R^5\over r_{\rm{peri}}^6}\,T(\eta,\omega/\Omega_{\rm{peri}},e),
\end{equation}
where $T$ is a dimensionless function of $\eta,$ $\omega/\Omega_{\rm{peri}},$ and $e$ (though $T$ becomes independent of $e$ as $e$ approaches unity). The exact form of $T(\eta,\omega/\Omega_{\rm{peri}},e)$ is provided in Appendix~\ref{sec:ApRealModel}.
Then,
\begin{equation}
|\Delta\hatP_1|={6\pi (\omega/\Omega_{\rm{peri}})\over (1-e)^{5/2}}\left({M'\over M}\right)
\left({M\over M_t}\right)^{\!5/3}\eta^{-5} T(\eta,\omega/\Omega_{\rm{peri}},e).
\label{eq:DeltahatP}
\end{equation}
In general, $|\Delta\hatP_1|$ falls off steeply with $\eta$. However, 
even when $\eta$ is large (weak tidal encounter), $|\Delta\hatP_1|$ can be significant 
for highly eccentric systems (with $1-e\ll 1$).

The map (\ref{eq:a-})-(\ref{eq:ak}) assumes that (i) energy transfer
occurs instantaneously at pericentre; (ii) at each pericentre passage, the change
in mode amplitude,
$a_{k-}-a_{k-1}$, is the same;
and (iii) the mode energy is always much less than the binding energy of the star.
The first condition is satisfied when $1-e\ll 1$. Eccentric systems that exhibit oscillatory behaviour (see Section \ref{NoDissipation}) easily satisfy this condition over many orbits. Chaotic systems (see Section \ref{NoDissipation}) can evolve through a larger range of mode energies and orbital eccentricities. For this condition to hold throughout evolution, they must begin with very large eccentricities.
The second condition requires that the 
pericentre distance remains constant, which in turn requires that 
the fractional change in orbital angular momentum, $\Delta L/L$, remain
small throughout orbital evolution. Using 
$\Delta L \sim \Delta E_{B}/\Omega_{\rm{peri}}$ as an estimate, we find that the condition $\Delta L/L\ll 1$ becomes
\begin{equation}
\frac{\Delta L}{L} \sim 
{1\over 2\sqrt{2}} \Delta\tildeE_{B} (1 - e) \ll 1.
\label{eq:delL}\end{equation}
Thus, for sufficiently eccentric orbits, $r_{\rm{peri}}$ is roughly constant even when the orbital energy changes by $\Delta \tildeE_{B}\sim 10$.
The third condition, $E_k\ll GM^2/R$, yields the expression
\begin{equation}
 \frac{\tildeE_k (1-e) }{2\eta}
\left({M'\over M}\right)\left(\frac{M_t}{M}\right)^{\!-1/3} 
\ll 1.
\label{eq:linear}\end{equation}
Again, this is easiest to satisfy for very eccentric orbits.
	
\begin{figure*}
	\begin{center}
		\includegraphics[width=5.5in]{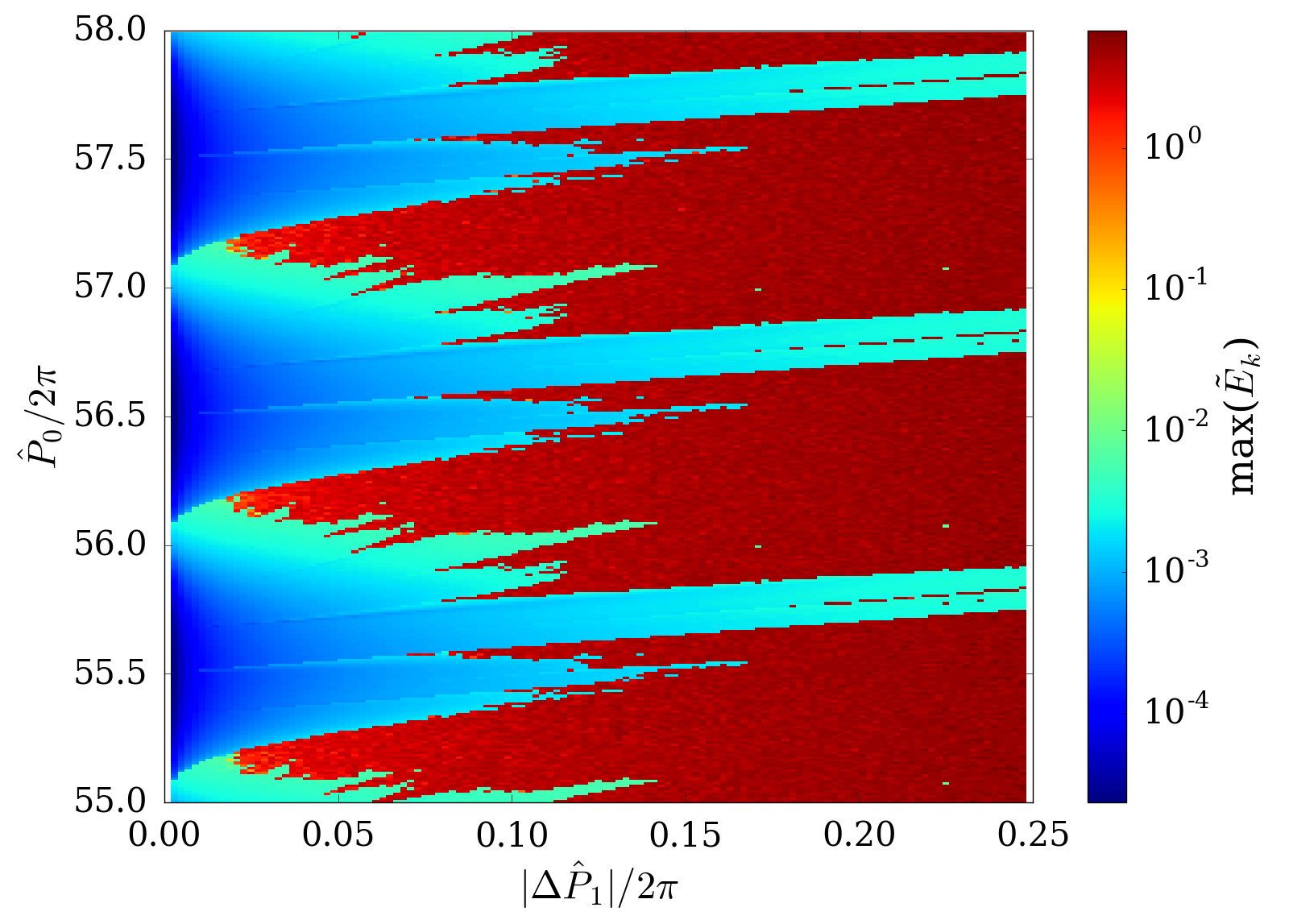}
		\caption{The maximum mode energy reached in $10^4$ orbits as a
			function of $|\Delta \hat{P}_1|/2\pi$ and $\hat{P}_0/2\pi$.  The
			energy is normalized to the initial orbital energy of the binary. In
			the dark blue regions, the mode exhibits low-energy oscillations. In
			the light green regions, the mode exhibits high-amplitude
			oscillations corresponding to a resonance. The red regions indicate
			chaotic mode evolution.}
		\label{fig:PhaseSpace}
	\end{center}
\end{figure*}
\section{Mode Energy Evolution without Dissipation}\label{NoDissipation}
\begin{figure*}
	\begin{center}
		\includegraphics[width=5.5in]{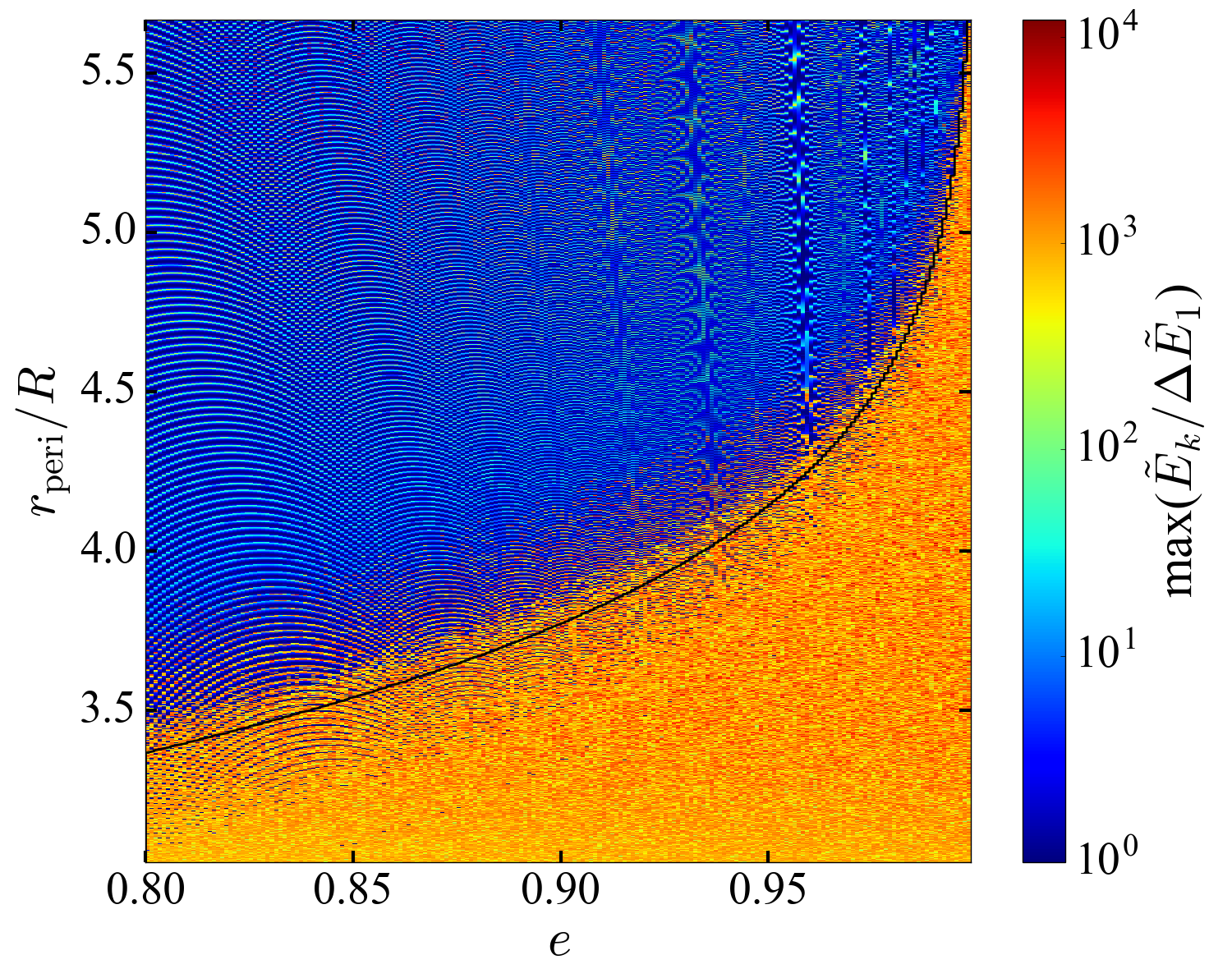}
		\caption{The maximum mode energy reached in $10^4$ orbits as a
			function of the pericentre distance, $r_{\rm peri}$, and $e$ for the $l=2$ f-mode of an $n=1.5$ polytrope in a binary with mass ratio $M'/M = 1$. The energy is normalized to the energy transferred in the first pericentre passage, $\Delta \tildeE_1$. In
			the dark blue region, the mode exhibits low-amplitude oscillations. The light blue ``fingerprint" ridges correspond to resonances. The yellow/orange region displays chaotic mode evolution. The black line indicates $|\Delta \hatP_1| = 1$. Note that, in this figure, the mode energy of the chaotic systems may not have attained the true ``theoretical" maximum [see equation~(\ref{eq:EmaxChaos})] in $10^4$ orbits; the energy may continue to climb to a large value if the system is allowed to continue evolving.}
		\label{fig:ChaosBoundary}
	\end{center}
\end{figure*}

We first study the dynamics of the ``eccentric orbit + mode" system without dissipation ($\hatgamma=0$).
The iterative map described in equations~(\ref{eq:a-})-(\ref{eq:ak})
displays a variety of behaviours depending on $\hat{P}_0$ and $|\Delta \hat{P}_1|$. 
We can gain insight into the evolution of the system
by recording $\tilde{E}_{\rm{max}}$, the maximum mode energy reached over many
orbits; this quantity reveals whether energy transfer to stellar
modes is relatively small or whether the orbit can change
substantially by transferring large amounts of energy. 
Figure \ref{fig:PhaseSpace} shows
$\tilde{E}_{\rm{max}}$ after $10^4$ orbits for systems with a range
of $\hat{P}_0$ and $|\Delta \hat{P}_1|$. Similarly, Fig.~\ref{fig:ChaosBoundary} displays $\tilde{E}_{\rm{max}}$ as a function of $r_{\rm peri}$ and $e$ for an $n=1.5$ polytrope stellar model in a binary with mass ratio $M'/M=1$. The relationship between the physical parameters $r_{\rm peri}$ and $e$ and the mapping parameters $\hatP_{0}$ and $|\Delta \hatP_1|$ is given by equations~(\ref{eq:hatP}) and (\ref{eq:DeltahatP}) (see Appendix~\ref{sec:ApRealModel} for more detail).

The system evolution has a complex dependence on $\hat{P}_0$ and
$|\Delta \hat{P}_1|$. In general, the mode energy exhibits oscillatory
behaviour for small $|\Delta \hat{P}_1|$ and chaotic growth for large
$|\Delta \hat{P}_1|$. However, Fig.~\ref{fig:PhaseSpace} shows exceptions to this trend. The figure also
suggests that the response to $\hat{P}_0$ is periodic and
the mode amplitude is larger in magnitude near resonances where the
orbital frequency is commensurate with the mode frequency. The map
displays three primary types of behaviours --- low-amplitude
oscillation, resonant oscillation, and chaotic evolution. Transitions
between the three regimes are complicated. However, within each regime,
$\tilde{E}_{\rm{max}}$ exhibits simple dependence on $\hat{P}_0$ and
$|\Delta \hat{P}_1|$. We now discuss the three types of behaviour in detail.
	
\subsection{Oscillatory Behaviour} \label{oscillate}
	
\begin{figure*}
\includegraphics[width = 5.5in]{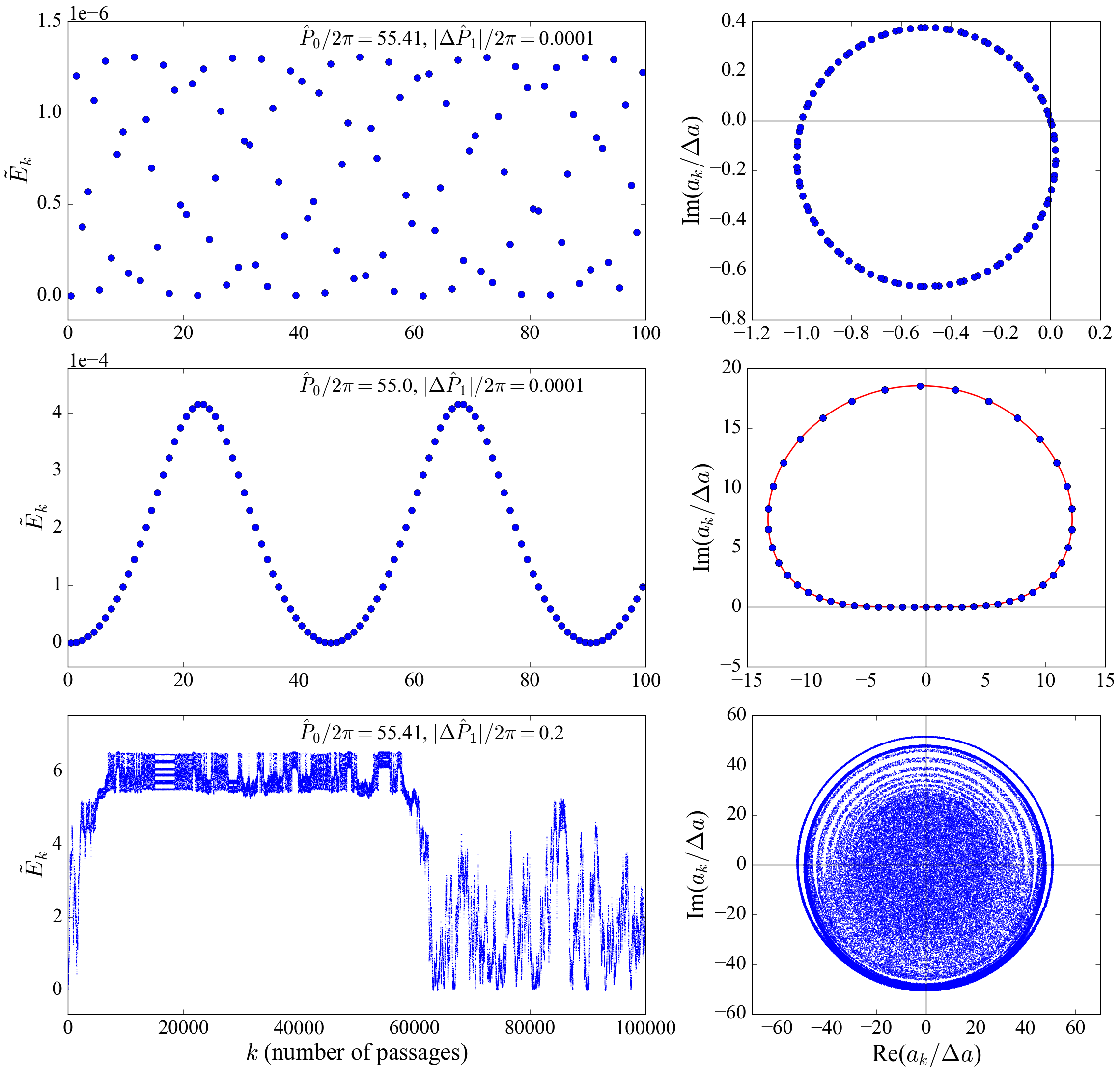}
\caption{Left Column: The evolution of the mode energy 
  over multiple pericentre passages. 
Right Column: The complex mode amplitude $a_k$ (normalized to $\Delta a$, the
change in mode amplitude during the first pericentre passage). From top to bottom, the three rows correspond to different types of behaviours ---
low-amplitude oscillations, resonant behaviour, and chaotic
evolution. The red line is from equation~(\ref{eq:akRes}) in
Appendix \ref{sec:ApRes}.}
\label{fig:3behaviours}
\end{figure*}
	
When 
$|\Delta\hatP_1|/(2\pi)\lesssim 0.05$
and $\hat{P}_0/(2\pi)$ is not close to an integer,
the mode exhibits low-amplitude oscillations, shown
in the top panels of Fig.~\ref{fig:3behaviours}. In this regime, the
orbital period is nearly constant ($\hatP_k \simeq \hatP_0$), and the
map from equations~(\ref{eq:a-})-(\ref{eq:ak}) can be written simply as
\begin{equation}
a_k \simeq (a_{k-1} + \Delta a)\, e^{-i\hatP_0}. \label{eq:OscMap} 
\end{equation}
This can be solved with the initial condition $a_0=0$, yielding
\begin{equation}
a_k \simeq \frac{\Delta a}{e^{i\hatP_0}-1}\left(1 - e^{-i\hatP_0 k}\right). \label{eq:OscSolution}
\end{equation}
Note that, in the complex plane, this solution has the form of a circle of radius 
$|1/(e^{i\hatP_0}-1)|$ centred on $1/(e^{i\hatP_0}-1)$, as shown in Fig.~\ref{fig:3behaviours} [a result previously seen in \citet{IP04}]. 
From equation~(\ref{eq:OscSolution}), the maximum mode energy in this regime is
\begin{equation}
\tilde{E}_{\rm{max}}\simeq \frac{2(\Delta a)^2}{1 - \cos{\hatP_0}}. \label{eq:EmaxOsc}
\end{equation}
This result demonstrates that our assumption of $\hatP_k \simeq
\hatP_0$ performs well when $(\Delta a)^2 \ll 1$ and $\hatP_0$ is not too close to an
integer multiple of $2\pi$. Under these conditions,
the mode energy remains of order $(\Delta a)^2=\Delta{\tilde E}_1$, the energy transfer
in the ``first'' pericentre passage.

\subsection{Resonance}\label{sec:resonance}
Figure \ref{fig:PhaseSpace} indicates that the stellar mode exhibits large-amplitude oscillations for $\hatP_0\simeq 2\pi N$ (with $N=$integer), 
i.e., when the orbital period $P_0$ is nearly an integer multiple of the mode period $2\pi/\omega$.
To understand this behaviour, we assume $\tildeE_k=|a_k|^2 \ll 1$, which holds true 
in the non-chaotic regime. With no dissipation, equation (\ref{eq:Kepler}) is replaced by 
\begin{equation}
\hatP_k \simeq \hatP_0\left(1-\frac{3}{2}|a_{k-}|^2\right)
=\hatP_0-|\Delta\hatP_1| |z_k|^2,
\end{equation}
where we have defined $z_k\equiv a_{k-}/\Delta a$ and used equation~(\ref{eq:defDelPhat}).
The map can then be written as 
\begin{equation}
z_{k+1}= 1+z_k\,e^{-i\hatP_k}\simeq
1+z_k\, e^{-i\hatP_0+i|\Delta\hatP_1| |z_k|^2}.
\label{eq:maplinear}
\end{equation}
Near a resonance, with $|\delta \hatP_0| = |\hatP_0 - 2\pi N|\ll 1$, the above map can be
further simplified to $z_{k+1}-z_k\simeq 1 +z_k (-i\delta\hatP_0 + i|\Delta P_1| |z_k|^2)$.
The maximum mode amplitude near resonance is determined by setting the non-linear term $|\Delta \hatP_1| |z_k|^3 \sim 1$, giving $|z_{\rm{res}}| \sim|\Delta\hatP_1|^{-1/3}$. The corresponding mode energy is
\begin{equation}
\tildeE_{\text{res}} \sim  {(\Delta a)^2 \over |\Delta \hatP_1|^{2/3}}
\sim  {|\Delta\hatP_1|^{1/3}\over \hatP_0}.
\label{eq:EmaxRes}
\end{equation}
Equation~(\ref{eq:EmaxRes}) is valid for $|\delta\hatP_0|\lesssim |\Delta\hatP_1|^{1/3}$,
and agrees with our numerical result (see Appendix \ref{sec:ApRes} for more details).

\subsection{Chaotic Growth}\label{sec:chaoticGrowth}

Chaotic growth of mode energy typically occurs when $|\Delta
\hat{P}_1| \gtrsim 1$, i.e., when enough energy is transferred in a
pericentre passage to change the orbital period and cause appreciable
phase shift of the mode.  In this regime, the mode amplitude fills a
circle in the complex plane after the binary evolves for many orbits,
as shown in the bottom panels of Fig.~\ref{fig:3behaviours}. 

We can verify that the dynamical behaviour of systems with a large $|\Delta \hatP_1|$ is
chaotic by examining the difference between a trajectory and its
shadow to estimate the Lyapunov exponent. The shadow trajectory is calculated
with a slightly different initial value $a_{0,{\rm shadow}}$, such that
$\delta a_0 \equiv |a_{0,\text{shadow}}| - |a_0|\ll 1$. We follow the 
evolution of $\delta a_k \equiv ||a_{k,\text{shadow}}| - |a_k||$. For
chaotic behaviour, we expect
\begin{equation}
\delta a_k \approx \delta a_0\, \text{e}^{\lambda k}, \label{eq:Lyapunov}
\end{equation}
where $\lambda$ is the Lyapunov exponent.

\begin{figure}
	\begin{center}
		\includegraphics[width=3.5in]{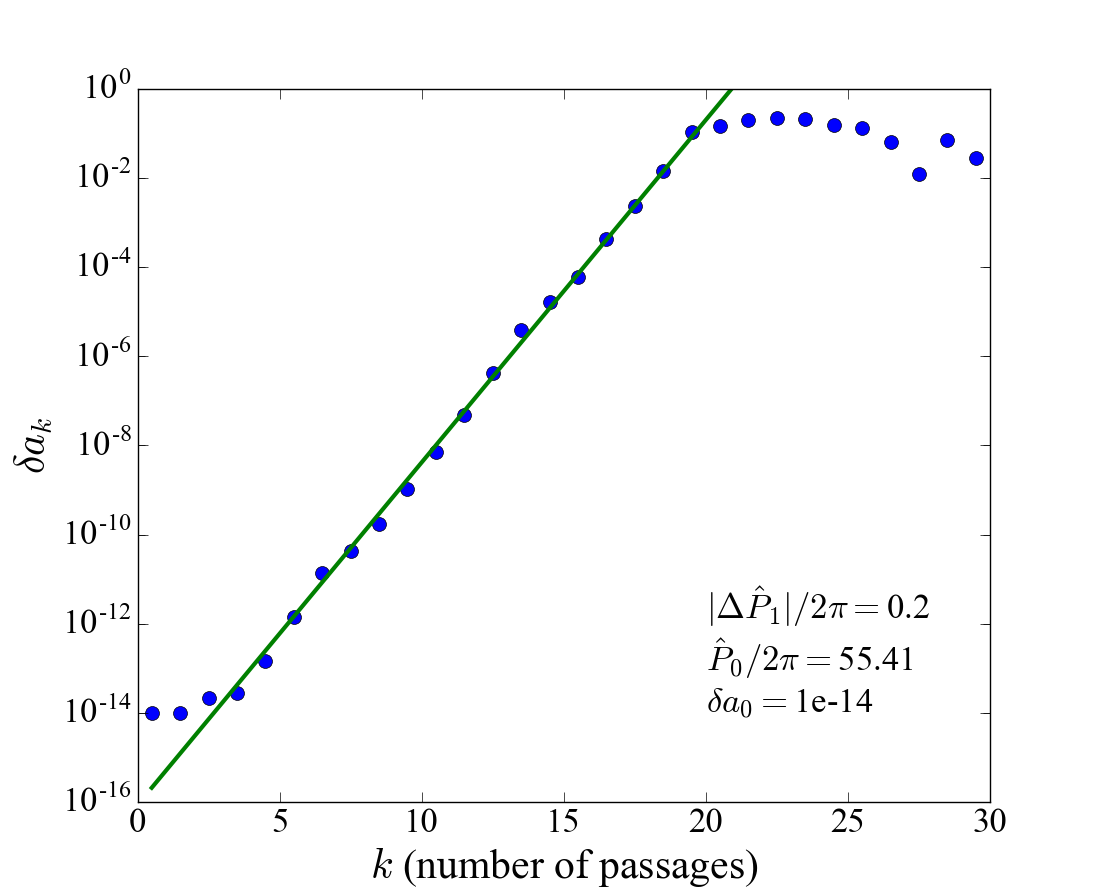}
		\caption{The blue dots show the difference between the mode amplitude
			of a trajectory and its shadow, $\delta a_k$, as a function of
			orbits. The green line shows a fit to the exponential rise with
			$\lambda \approx 1.77$, where $\lambda$ is defined in
			equation~(\ref{eq:Lyapunov}). For $k \gtrsim 20$, $\delta a_k$ saturates
			around 0.1. }
		\label{fig:Lyapunov}
	\end{center}
\end{figure}

Figure \ref{fig:Lyapunov} suggests that systems  with $|\Delta \hatP_1| \sim 1$ indeed undergo chaotic evolution, with $\delta a_k$ growing exponentially (but eventually saturating
when $\delta a_k \sim 0.1$). For the system depicted in Fig.~\ref{fig:Lyapunov}, $\lambda \approx 1.77$. 
The exact value of $\lambda$ can change slightly with the parameters $\hatP_0$ and $|\Delta \hatP_1|$. Similar Lyapunov calculations were preformed in \citet{Mardling95a,Mardling95b} to determine numerically the boundary for chaotic behaviour in the $r_{\rm peri} - e$ plane [e.g. Fig. 13 in \citet{Mardling95a}, which qualitatively agrees with our Fig.~\ref{fig:ChaosBoundary}]. The condition $|\Delta \hatP_1| \gtrsim 1 $ for chaotic behaviour was first identified by \citet{IP04}.

\begin{figure*}
	\begin{center}
		\includegraphics[width=5in]{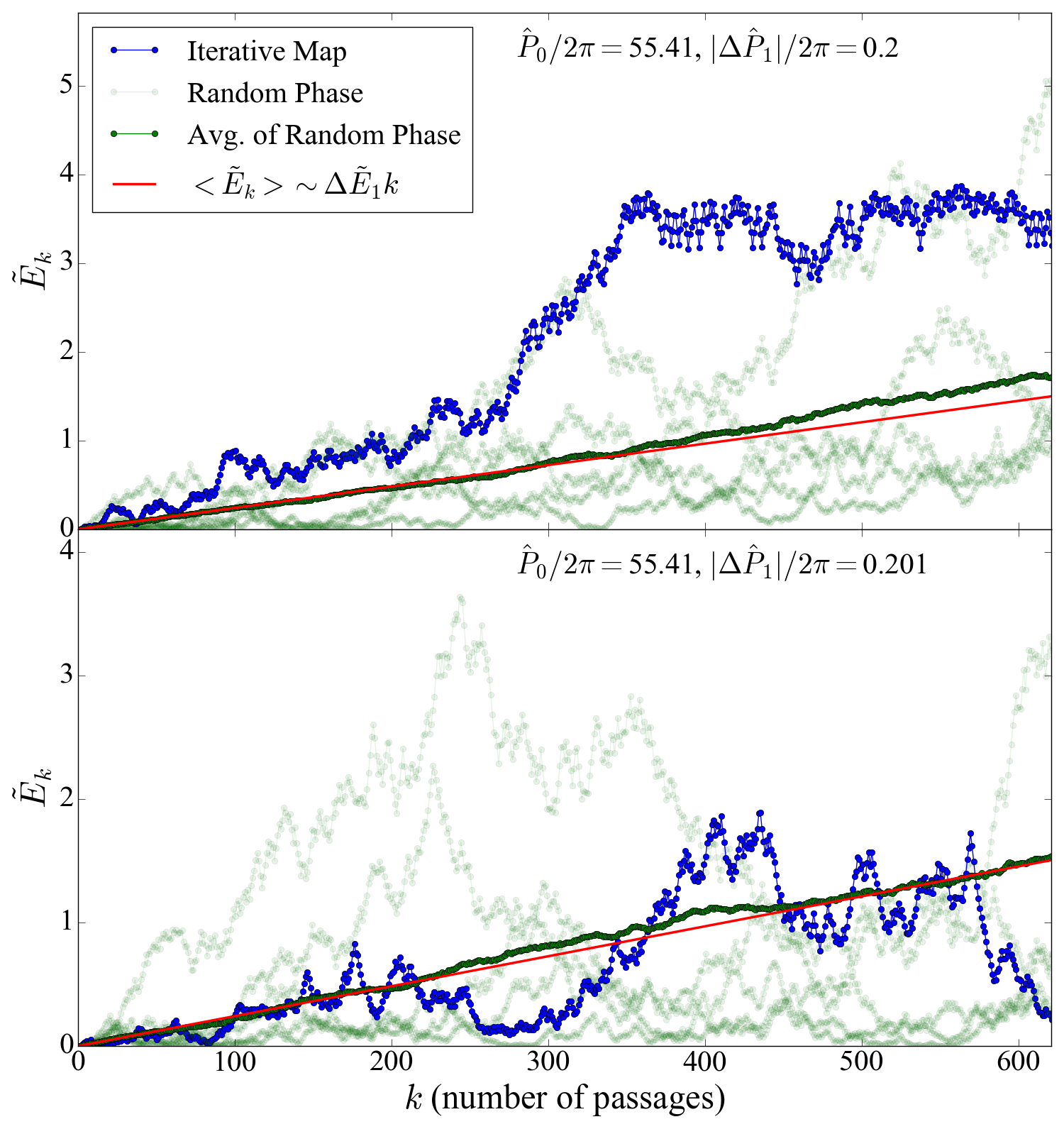}
		\caption{The mode energy evolution for two slightly different values
			of $|\Delta \hatP_1|$. The blue lines are the results from the iterative
			map; the light green lines show examples of the random phase model [where
			$\hatP_k$ in equation~(\ref{eq:hatpk}) takes on random values between 0 and
			$2\pi$]; the dark green is an average of 100 iterations of the random phase model; and the red lines show the expected average growth of the
			mode energy in the diffusion model.}
		\label{fig:RandomPhase}
	\end{center}
\end{figure*}

While the Lyapunov saturation of $\delta a_k$ occurs after 10's of orbits, the mode energy can continue to climb over much longer timescales (as in the bottom panel of Fig.~\ref{fig:3behaviours}).
In the absence of dissipation, the mode amplitude map is simply 
\begin{equation}
a_k=(a_{k-1}+\Delta a)\,e^{-i\hatP_k}, \quad
{\rm with}~~\hatP_k=\hatP_0 \left(1+|a_k|^2\right)^{-3/2}.
\label{eq:hatpk}\end{equation}
When the change in orbital period between
pericentre passages, $\Delta\hatP_k=\hatP_k-\hatP_{k-1}$, is much larger than unity,
$\hatP_k$ approximately takes on random phases. This random-phase model, previously studied in \citet{Mardling95a, IP04}, captures 
the key features of mode growth in the chaotic regime (see Fig. \ref{fig:RandomPhase}). 
The mode energy after the $k$-th passages can be written as
\begin{equation}
\tilde E_k=\sum_{j=1}^k\Delta\tilde E_j=
\sum_{j=1}^k\left[(\Delta a)^2+2(\Delta a){\rm Re}(a_{j-1})\right].
\end{equation}
If we {\it assume} that $a_{j-1}$ exhibits random phases, then 
\begin{equation}
\langle \tilde E_k\rangle \sim
(\Delta a)^2 k, \label{eq:AvgChaoticGrowth}
\end{equation}
a result previously obtained by \citet{Mardling95a} and \citet{IP04}. This provides a crude description of the chaotic mode growth, shown in Fig. \ref{fig:RandomPhase}.

Although $\tilde{E}_{\text{max}}$ can become very large,
Fig.~\ref{fig:3behaviours} suggests that the mode energy cannot exceed
a maximum value, a feature not captured by the random-phase model, but previously seen in some examples of numerical integrations of chaotic mode evolution \citep{Mardling95b}. 
This can be understood from the fact that as the mode energy increases, the 
the range of possible $\Delta \hatP_k$ decreases. Indeed, from equation~(\ref{eq:hatpk})
we find 
\begin{equation}
\Delta \hatP_k \equiv \hatP_k - \hatP_{k-1} \simeq 
-3\Delta a \hatP_0 \frac{\real (a_{k-1})}{(1 +|a_{k-1}|^2)^{5/2}} \label{eq:Pdiff}.
\end{equation}
Setting $\Delta\hatP_k\sim 1$ leads to a maximum mode energy 
\begin{equation}
\tilde{E}_{\text{max}}=\bigl(|a_k|^2\bigr)_{\rm max} \sim (\hatP_0 \Delta a)^{1/2}\sim 
\left(|\Delta\hatP_1|\hatP_0\right)^{1/4}. \label{eq:EmaxChaos}
\end{equation}
More discussion on the maximum mode energy in the chaotic regime 
can be found in Appendix \ref{sec:ApC}. 
Note that $\tildeE_{\text{max}}$ of order a few can be easily reached 
for a large range of $\hatP_0$ and $|\Delta \hatP_1|$ (see Fig. \ref{fig:3behaviours}). Such a large mode 
energy implies order unity change in the semi-major axis of the orbit,
but for $1-e\ll 1$ it does not necessarily violate the requirements needed 
for the validity of the iterative map [see equations~(\ref{eq:delL})-(\ref{eq:linear})].

	
\section{Mode Energy Evolution with Dissipation} \label{Dissipation}

We now consider the effect of dissipation on the evolution of the
system. In the presence of mode damping ($\gamma\neq 0$), energy is
preferentially transferred from the orbit to the stellar mode which then
dissipates, causing long-term orbital decay. In the extreme case when the mode 
damping time $t_{\rm damp}=\gamma^{-1}$ is shorter than the orbital period $P$, 
the energy transfer in each pericentre passage $\Delta E_1$ is dissipated, and 
the orbital energy $E_B$ simply decays according to 
\begin{equation}
{dE_B\over dt}\simeq -{\Delta  E_1\over P},
\quad ({\rm for} ~t_{\rm damp}\lesssim P).
\end{equation}
Below, we will consider the more realistic situation
of $t_{\rm damp}\gg P$.

\subsection{Quasi-Steady State}
\begin{figure}
\begin{center}
\includegraphics[width=\columnwidth]{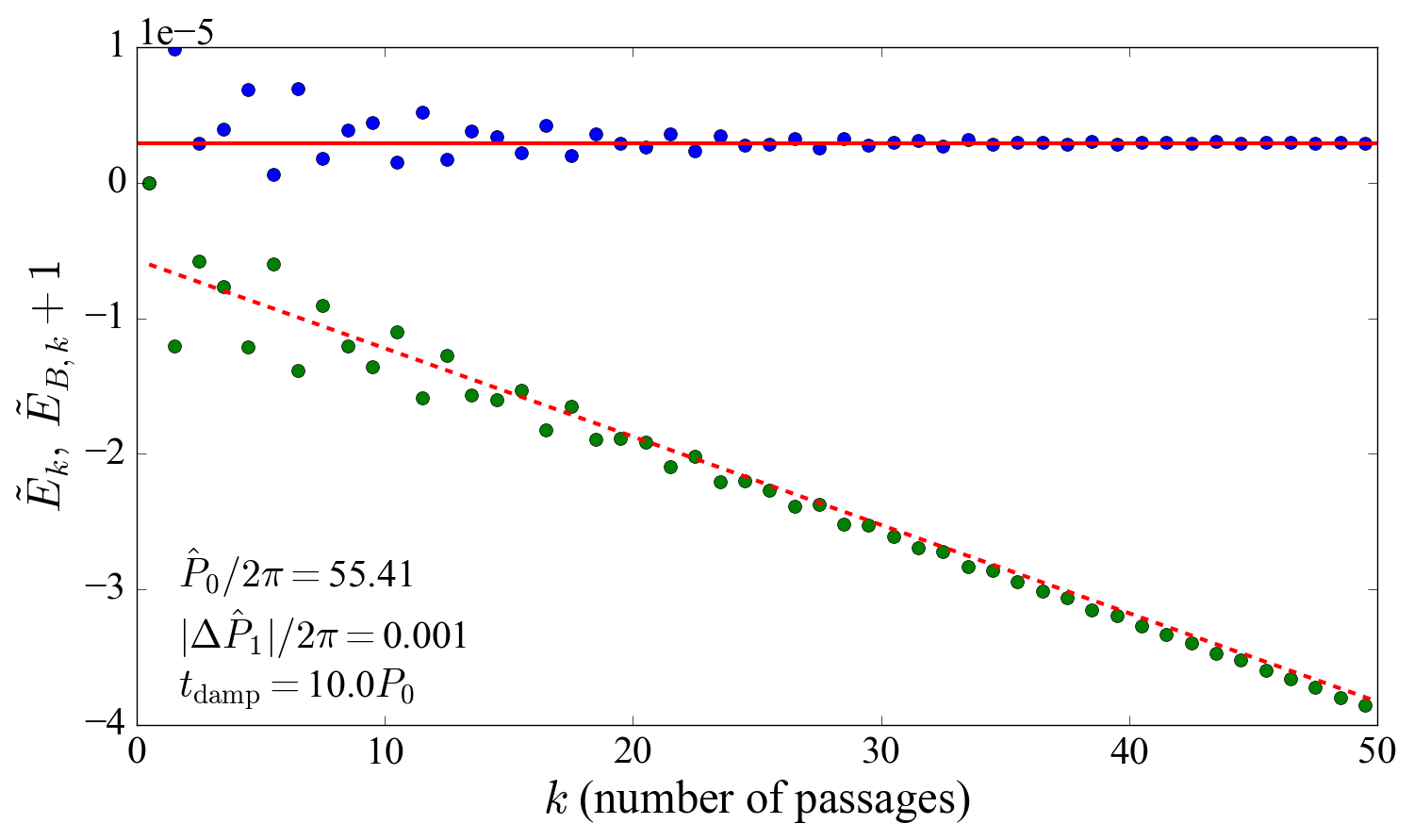}
\caption{The stellar mode energy (blue dots) and the 
  orbital energy (green dots; shifted for comparison) over $50$ pericentre
  passages. Energy is normalized to the initial orbital energy of the
  binary. The solid red line shows the steady-state mode energy
  given by equation~(\ref{eq:QSS}), and the dashed red line shows the
  orbital decay rate from equation~(\ref{eq:eTransRateOsc}).}
\label{fig:EnergyOscDamp}
\end{center}
\end{figure}

Consider a system with $|\Delta \hatP_1|\ll 1$ and an orbital period
that is far from resonance. The mode energy will stay around $\Delta E_1$,
and can attain a quasi-steady state after a few damping times 
(see Fig.~\ref{fig:EnergyOscDamp}).
Indeed, since the orbital period $P$ remains roughly constant over multiple damping times,
the map simplifies to 
\begin{equation}
a_k \simeq  (a_{k-1} + \Delta a)\, e^{-(i + \hat{\gamma})\hatP}. \label{eq:OscMapDamp}
\end{equation}
Assuming $a_0 = 0$, we find
\begin{equation}
a_k \simeq \frac{\Delta a}{\text{e}^{(\text{i} + \hat{\gamma})\hat{P}}-1}
\left[1 - \text{e}^{-(\text{i} + \hat{\gamma})\hat{P} k}\right]. \label{eq:OscSolutionDamp}
\end{equation}
Clearly, the mode amplitude approaches a constant value after a few damping times 
($kP\gg \gamma^{-1}$), and the mode energy reaches the steady-state value \citep{Lai97}:
\begin{equation}
\tilde{E}_{\rm{ss}} \simeq \frac{\Delta \tildeE_1 \,
  e^{-\hat{\gamma}\tilde{P}}}
{2(\cosh \hat{\gamma}\hat{P} - \cos\hat{P})}
\simeq \frac{\Delta \tildeE_1}{4 \sin^2 (\hatP/2) + (\hatgamma \hatP)^2}, \label{eq:QSS}
\end{equation}
where the second equality assumes $\hat{\gamma}\hatP=\gamma P \ll 1$, or $t_{\rm{damp}}\gg P$.
The steady-state energy is of order $\Delta \tildeE_1$, provided that the system is not near a resonance.
In the steady state, the star dissipates all the ``additional'' energy gained at each pericentre passage,
and thus the orbital energy decays according to 
\begin{align}
{d{\tilde{E}}_B\over dt}\simeq -2\gamma \tilde{E}_{\text{ss}}~~~{\rm or}~~~
{dE_B\over dt}\simeq -{2{E}_{\text{ss}}\over t_{\rm damp}}, && \quad ({\rm for} ~t_{\rm damp}\gg P) .
\label{eq:eTransRateOsc}\end{align}

\subsection{Passing Through Resonances}\label{ResonanceDissipation}
\begin{figure}
\begin{center}
\includegraphics[width=\columnwidth]{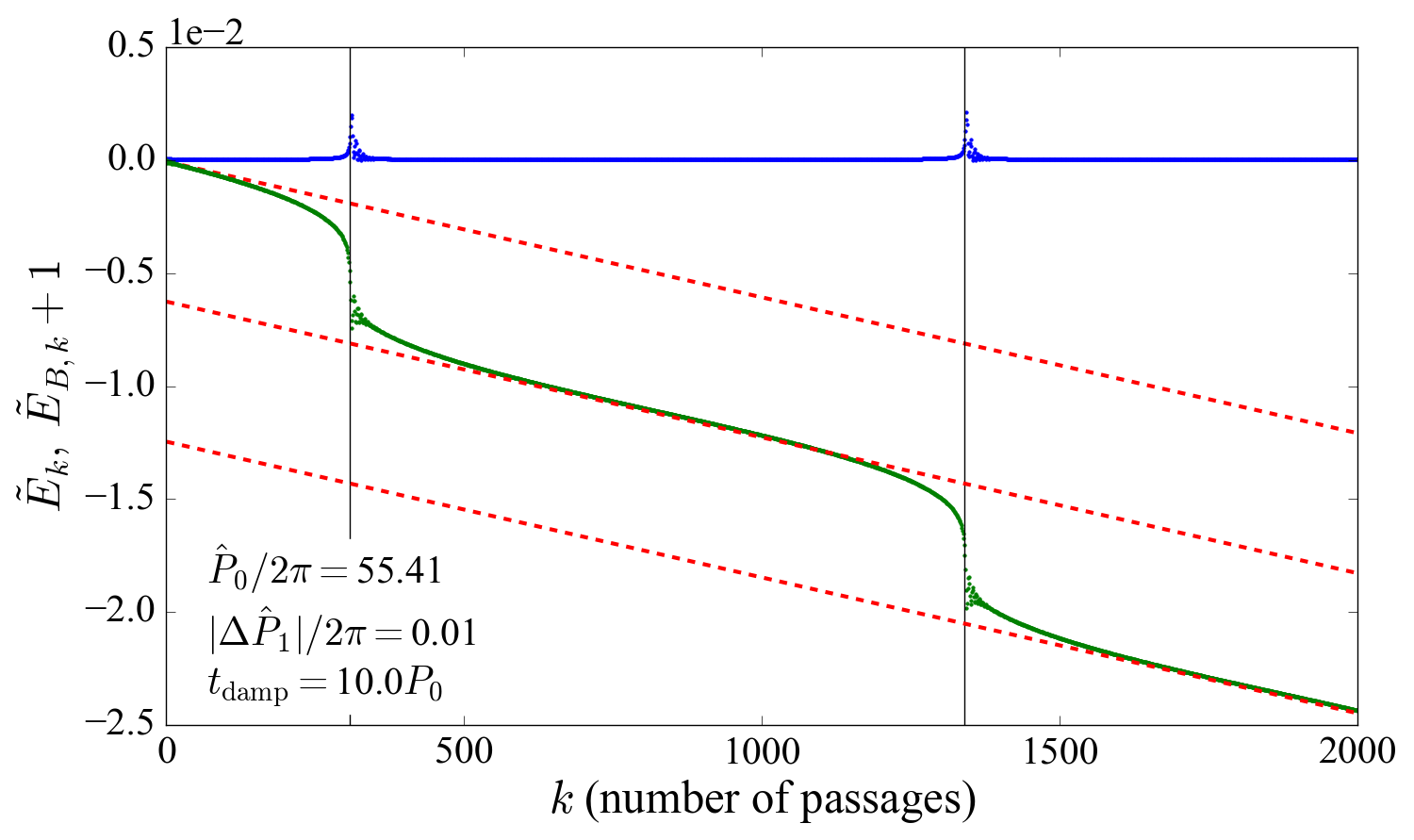}
\caption{The mode energy (in blue) and the orbital energy (in green) over many pericentre passages. Energy is
  normalized to the initial orbital energy of the binary. The analytic 
  orbital decay rate for the quasi-steady state is plotted in dashed 
  red. The red lines are spaced by $5.46 \tildeE_{\rm res}$ 
with $\tildeE_{\rm{res}}$ given by equation~(\ref{eq:EmaxRes}).}
\label{fig:EnergyMultiRes}
\end{center}
\end{figure}

As the binary orbit experiences quasi-steady decay, it will encounter 
resonances with the stellar mode ($\hatP/2\pi=$integer), during which rapid orbital decay
occurs (see Fig.~\ref{fig:EnergyMultiRes}).

The change in orbital energy when a system moves through a resonance
depends on how the resonance time $t_{\rm res}$ (the timescale for the
mode energy of a system near resonance to reach
$\tildeE_{\text{res}}$) compares with $t_{\text{damp}}$. In most
likely situations, the resonance time $t_{\rm res}\sim
P/|\Delta\hatP_1|^{1/3}$ (see Appendix \ref{sec:ApResTimescale}) is much
shorter than $t_{\rm damp}$, so the orbital energy is quickly
transferred to the stellar mode as the system approaches the
resonance, and the mode energy reaches the maximum resonance value given by equation~(\ref{eq:EmaxRes}). All of this energy is dissipated within a few $t_{\rm damp}$, 
resulting in a net change in the orbital energy during the resonance
$\Delta{\tilde E}_{B,{\rm res}}\simeq {\tilde E}_{\rm res}\sim |\Delta\hatP_1|^{1/3}/\hatP$.
By comparison, the quasi-steady orbital energy change between adjacent 
resonances [from $\hatP=2\pi N$ to $\hatP=2\pi (N-1)$] is (assuming $N\gg 1$)
\begin{equation}
\Delta \tildeE_{B,\text{ss}} \simeq \frac{4\pi}{3 \hatP}.
\label{eq:QSSRes}
\end{equation}
Thus $\Delta{\tilde E}_{B,{\rm res}}/\Delta \tildeE_{B,\text{ss}} \sim 0.2 |\Delta \hatP_1|^{1/3}$.
In practice, systems that evolve into resonance rather than starting in resonance will 
reach a maximum mode energy of a few times equation~(\ref{eq:EmaxRes}), so
$\Delta{\tilde E}_{B,{\rm res}}$ can be comparable to $\Delta \tildeE_{B,\text{ss}}$.

\subsection{Tamed Chaos}\label{sec:tamedChaos}
\begin{figure*}
\begin{center}
\includegraphics[height=5in]{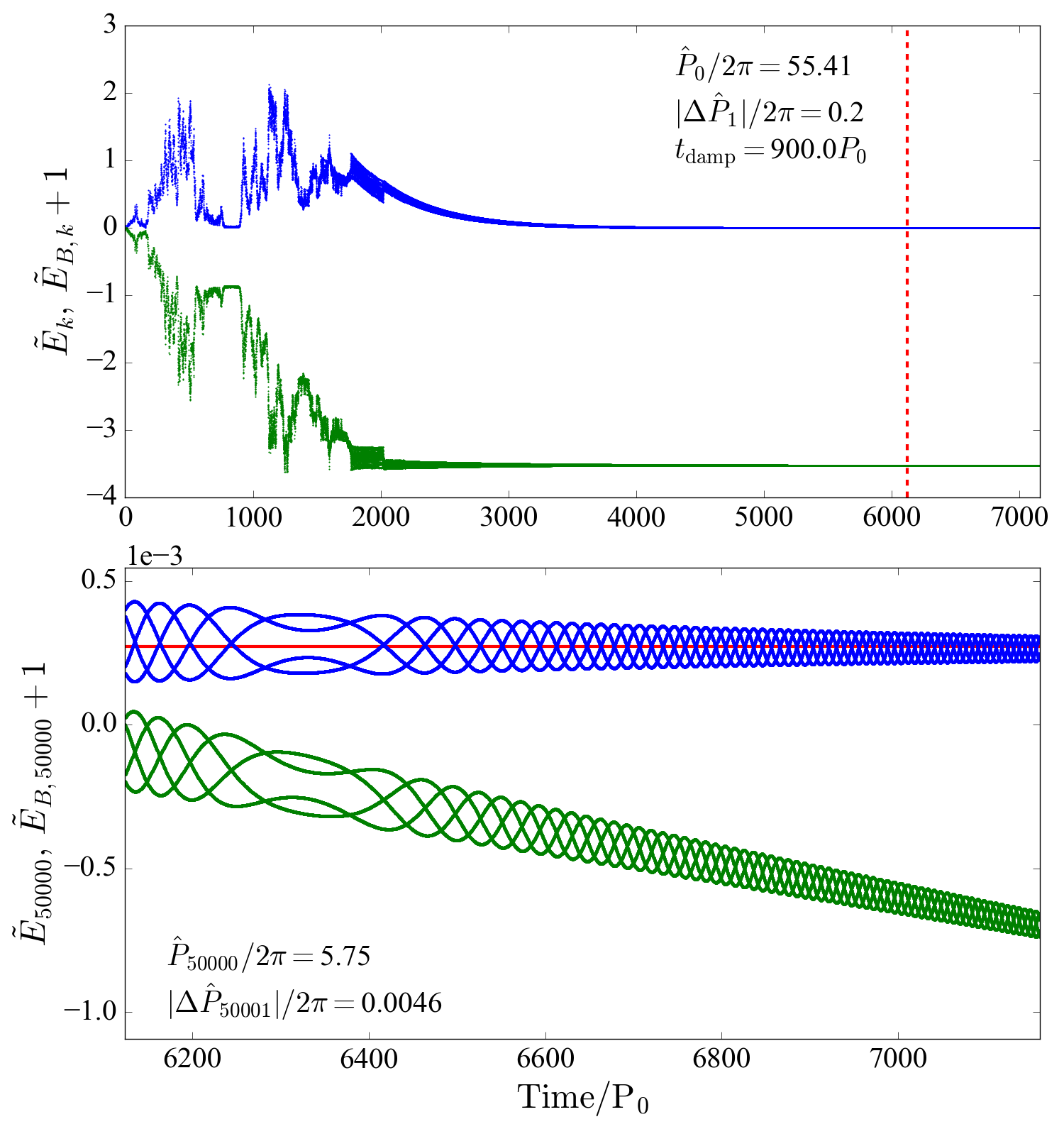}
\caption{The mode energy (in blue) and the binary orbital energy (in green) over many pericentre passages. In the upper panel, the energy is
  normalized to the initial orbital energy of the binary. The mode
  energy undergoes chaotic evolution, and damps to a quasi-steady sate
  after a few $t_{\text{damp}}$. The lower panel shows the later phase of the evolution
  (to the right of the dashed red line in the upper panel), with the energy 
re-normalized to the energy of the binary after 50,000 orbits. The predicted quasi-steady
  state mode energy is shown as a solid red line. The values of $\hatP$ and $|\Delta\hatP|$
  at 50,000 orbits are indicated. Note that because of the significant orbital decay during chaotic evolution, we use time (in units of $P_0$) rather than the number of orbits for the $x$-axis.}
\label{fig:EnergyTamedChaos}
\end{center}
\end{figure*}

In the presence of dissipation, even systems that experience chaotic
mode growth eventually settle into a quasi-steady state.  Figure
\ref{fig:EnergyTamedChaos} depicts an example. We see that initially
the mode energy increases rapidly, accompanied by a large decrease in
the orbital energy. This behaviour has been seen in numerical integrations of forced stellar oscillations and orbital evolution by \citet{Mardling95b}, where the orbital eccentricity quickly decreases to a value dictated by the ``chaos boundary" before settling into a state of gradual decay. With our exact ``dissipative" map, we can predict the steady-state mode energy and orbital decay rate that emerge after a period of chaotic evolution. For systems with relatively large damping rates,
the mode energy may not reach the full ``chaotic maximum'' given by
equation~(\ref{eq:EmaxChaos}). However, 
for systems with relatively small damping rates, the full maximum energy is attainable. 
In either case, the mode energy ultimately decays to a quasi-steady value of order
$\Delta E_1$ after a timescale of $\sim t_{\rm damp} \ln(E_{\rm  max}/\Delta E_1)$. 
The evolution of the system in quasi-steady state is well described by 
equations~(\ref{eq:QSS})-(\ref{eq:eTransRateOsc}).

We can understand how an initially chaotic system (with
$|\Delta\hatP_1|\sim 1$) is brought into the ``regular'' regime by
renormalizing various quantities to their ``post-chaotic'' values (see
the lower panel of Fig.~\ref{fig:EnergyTamedChaos}). Recall that the 
key parameter that determines the dynamical state of the system is 
$|\Delta\hatP|=\omega |\Delta P|$, with $\Delta P$ the change in the orbital period 
in a the first pericentre passage (i.e. when there is no prior oscillation). Since 
$|\Delta P|/P\simeq 3\Delta E/(2|E_B|)$ and $\Delta E$ is independent of the semi-major axis
$a$ (it depends only $r_{\rm{peri}}$, which is almost unchanged), we find
$|\Delta\hatP|\propto a^{5/2}\propto (1-e)^{-5/2}$ [for $r_{\rm{peri}}=$ constant; see equation~(\ref{eq:DeltahatP})].
Thus, after significant orbital decay (with $a$ decreased by a factor of a few),
$|\Delta\hatP|$ is reduced to a ``non-chaotic'' value, and the system settles into 
the regular quasi-steady state.

We can approximate the orbital parameters of a ``tamed" chaotic system that has reached quasi-steady state from the evolution of the orbital energy, $\tildeE_{\rm B} $. Our map assumes that angular momentum is conserved as the orbit evolves. Given this constraint, the orbital eccentricity just before the $(k+1)$-th pericentre passage is
	\begin{equation}
	e_k = \left[1-|\tildeE_{B,k}|(1-e_0^2)\right]^{1/2}.
	\end{equation}
As an example, a system with initial eccentricity $e_0=0.99$ that settles to a quasi-steady state orbital energy $\tildeE_{\rm B} \approx -5$ would retain an eccentricity of $e \approx 0.95$. Note that less eccentric binaries (even $e=0.9$) can circularize substantially over the course of chaotic evolution and strain the assumptions of our map (see Section~\ref{mapping}).

Our model assumes linear mode damping. In reality, modes that are
excited to high amplitudes may experience non-linear damping.  This
will likely make the system evolve to the quasi-steady state more
quickly. Other than this change of timescale, we expect that the
various dynamical features revealed in our model remain valid.
We note that a rapidly heated star/planet may undergo significant
structural change depending on where heat is deposited. This may alter
the frequencies of stellar modes. Our current model does not account
for such feedback.
	
\section{Systems with Multiple Modes}\label{MultipleModes}

Our analysis can be easily generalized to systems with multiple modes (labelled by
the index $\alpha$). The total energy in modes just before the $(k+1)$-th passage is ${\tilde E}_{k} = \sum_\alpha |a_{\alpha,k}|^2$.
During a pericentre passage, the amplitude of each mode changes by
$\Delta a_\alpha$. The total energy transferred to stellar modes in the $k$-th passage is
\begin{equation}
\Delta\tildeE_{k} = \sum_\alpha \left[|a_{\alpha,k-1}+\Delta a_\alpha|^2-|a_{\alpha,k-1}|^2\right].
\end{equation}
As before, the orbital energy after the $k$-th passage is given by $\tildeE_{B,k} = \tildeE_{B,0} - \sum_{j=1}^k \Delta\tildeE_{j}$.
The relationship between the orbital energy and the period is given by
equation~(\ref{eq:Kepler}). The mode amplitude of each mode just before the $(k+1)$-th passage is
\begin{equation}
a_{\alpha,k} = [a_{\alpha,k-1} + \Delta a_\alpha]\,\text{e}^{-(\text{i}+\hatgamma_\alpha)\hatP_{\alpha,k}}
\end{equation}
where $\hatP_{\alpha,k} \equiv \omega_\alpha P_k$. Similarly, we define $|\Delta\hatP_{\alpha,1}| 
\equiv \omega_\alpha|\Delta P_{1}|$. The evolution of the system is completely determined by
$\hatP_{\alpha,0}$, $|\Delta \hatP_{\alpha,1}|$ and $\hatgamma_\alpha = \gamma_\alpha/\omega_\alpha$.
	
\begin{figure*}
	\begin{center}
		\includegraphics[width=4.5in]{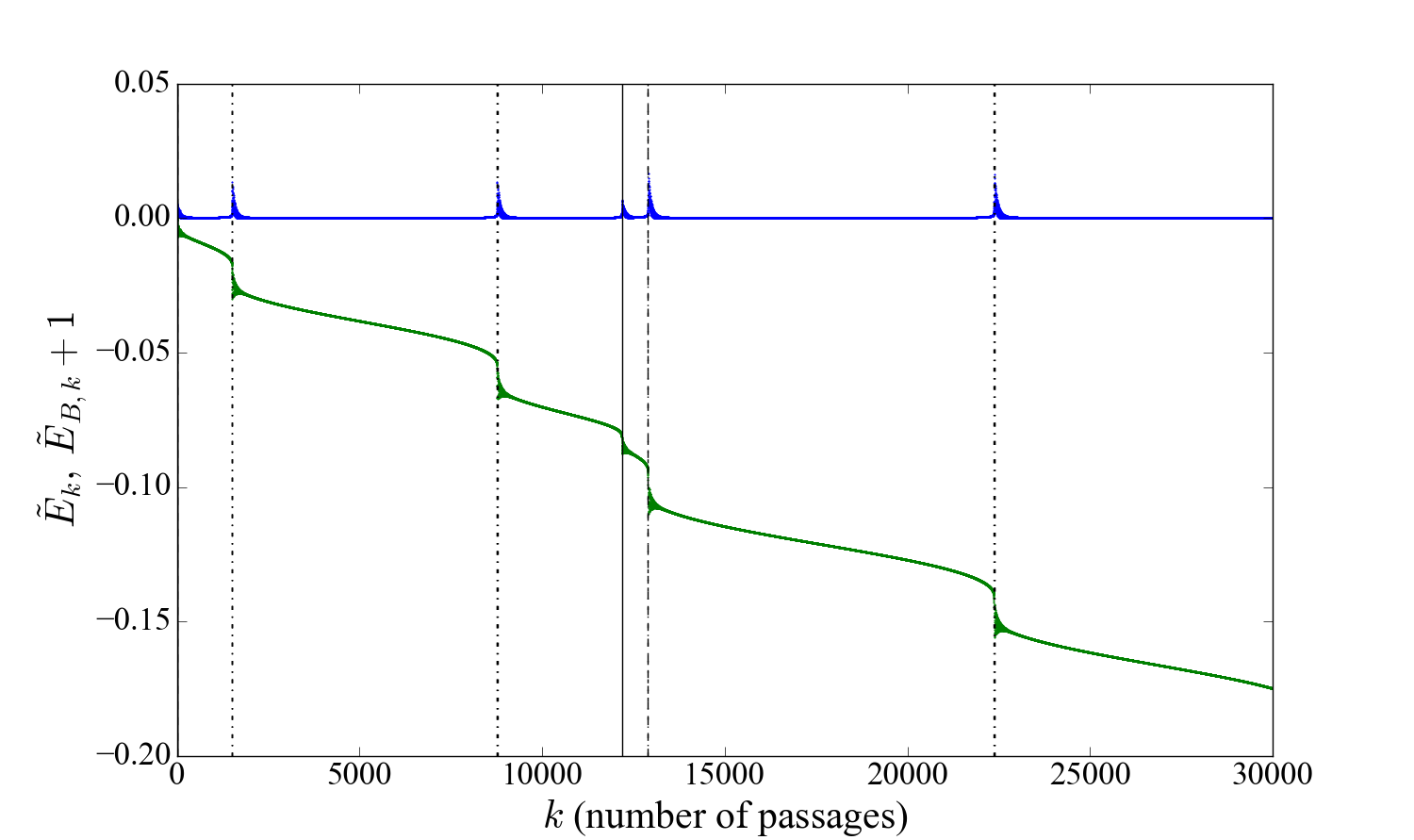}
		\caption{The total mode energy and orbital energy of a system with three
			modes that evolves through multiple resonances. 
			The properties of the modes are discussed in detail in Appendix \ref{sec:ApE}. 
			For this system, the parameters are $\hatP_{\alpha,0}/2\pi = 123,
			56.6, 50.4$ and $|\Delta \hatP_{\alpha,1}|/2\pi = 0.1, 0.05 , 0.04$. 
			The mode damping times are all of order $t_{\rm{damp}}\sim 100P_0$. 
			Resonances for different modes are shown with vertical solid, dashed, and
			dot-dashed lines.}
		\label{fig:MultiModeResonance}
	\end{center}
\end{figure*}

\begin{figure*}
	\begin{center}
		\includegraphics[width=4.1in]{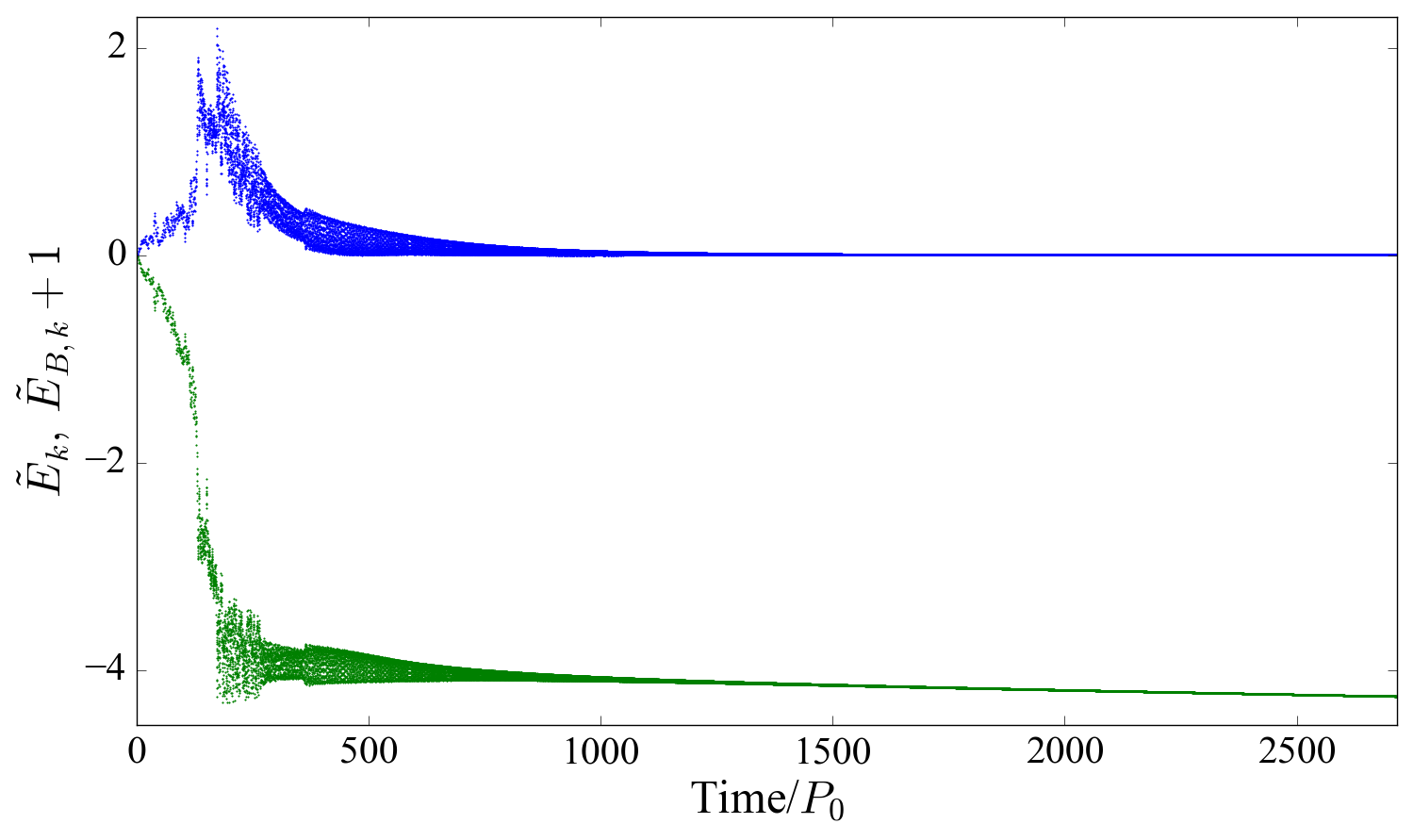}
		\caption{The total mode energy and orbital energy of a system with three
			modes that undergoes initial chaotic growth and eventually damps to a quasi-steady state. 
			The properties of the modes are discussed in detail in Appendix \ref{sec:ApE}. 
			For this system, the parameters are
			$\hatP_{\alpha,0}/2\pi= 123, 137, 113$ and $|\Delta \hatP_{\alpha,1}|/2\pi =2.3, 2.5, 2.1$. 
			The mode damping times are all of order $t_{\rm{damp}}\sim 100P_0$.}
		\label{fig:MultiModeChaos}
	\end{center}
\end{figure*}

In general, systems with multiple modes exhibit the same types of
behaviours seen in the single-mode system. Systems with small $|\Delta\hatP_{\alpha,1}|$ 
pass through multiple resonances over many orbits. For
systems with many modes, resonances play a significant role
in the orbital evolution, as shown in Fig. \ref{fig:MultiModeResonance}. Multi-mode systems also exhibit chaotic growth
that damps into a quasi-steady state (see Fig.~\ref{fig:MultiModeChaos}). 

\begin{figure*}
\includegraphics[width=3.45in]{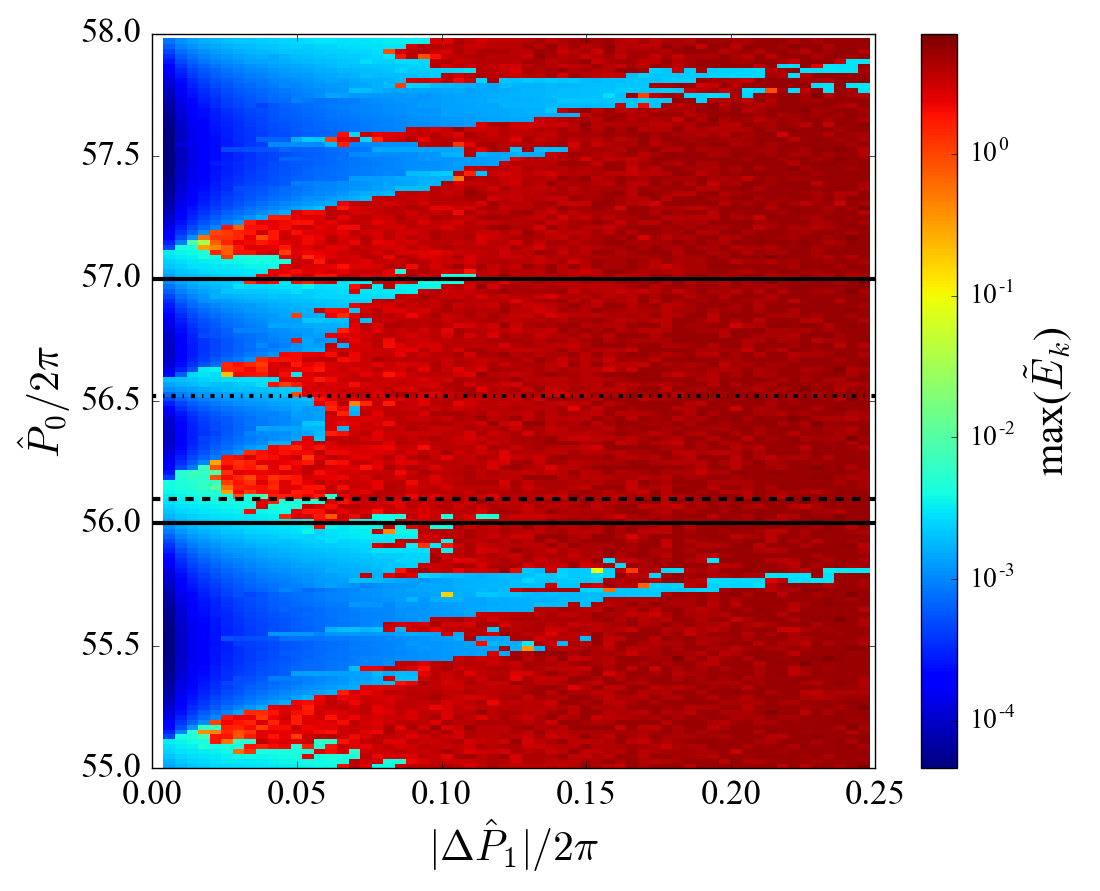}
\includegraphics[width=3.45in]{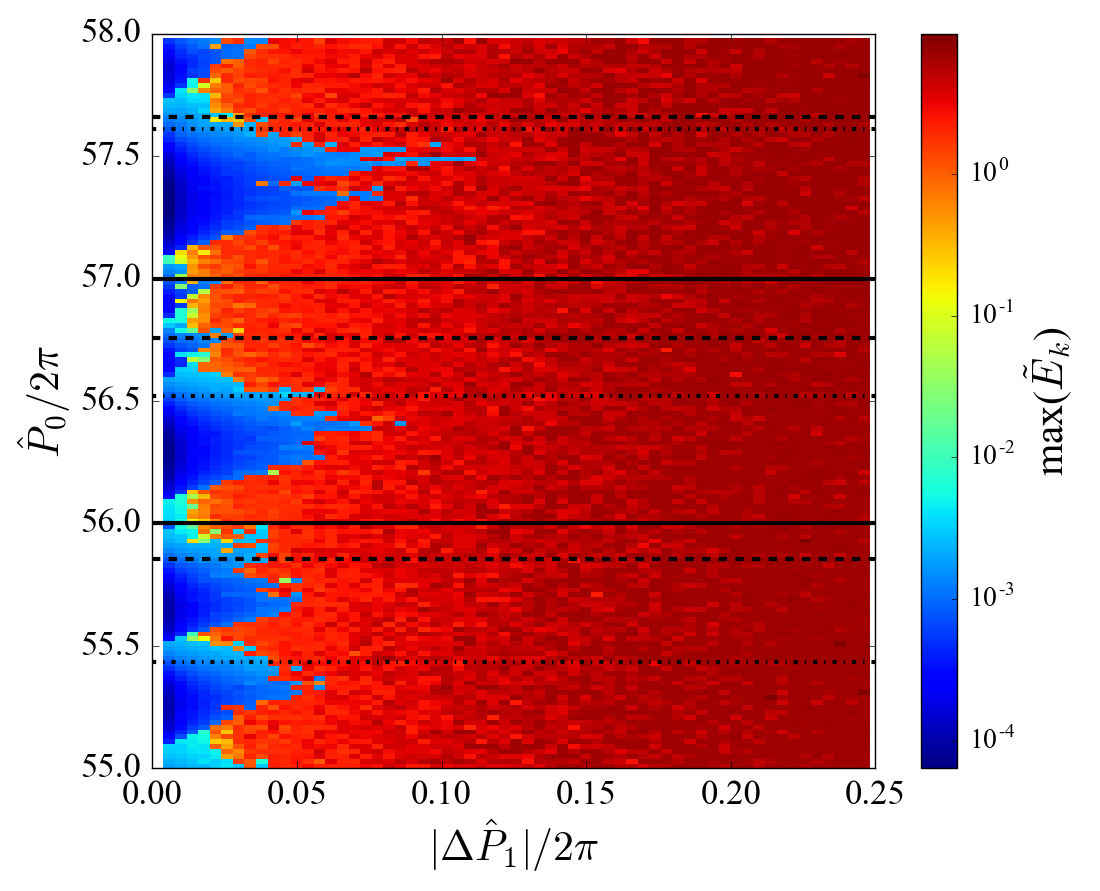}
\caption{The maximum mode energy (summed over all modes) reached in
  10,000 orbits as a function of $\omega_1|\Delta P_1|/2\pi$ and
  $\omega_1 P_0/2\pi$. In the dark blue regions, the modes exhibit
  low-energy oscillations, while in the dark red regions the modes grow chaotically to large amplitudes. The left panel shows a system with one
  dominant mode. The frequency ratios are $\omega_2/\omega_1 = 0.41$
  and $\omega_3/\omega_1 = 0.46$. The energy ratios are $\Delta \tildeE_{2,1}/\Delta \tildeE_{1,1} = 0.04$ and $\Delta \tildeE_{3,1}/\Delta \tildeE_{1,1} =
  0.03.$ The dashed line corresponds to $\omega_2 \hatP_0/2 \pi = 23$,
  the dot-dashed line to $\omega_3 \hatP_0/2 \pi = 26$, and the solid
  lines to resonances for $\omega_1$.  The right panel shows a system
  where all three modes are excited to similar energies. The
  frequency ratios are $\omega_2/\omega_1 = 1.11$ and
  $\omega_3/\omega_1 = 0.92$. The energy ratios are $\Delta \tildeE_{2,1}/\Delta \tildeE_{1,1} = 0.94$ and $\Delta \tildeE_{3,1}/\Delta \tildeE_{1,1} =
  0.58$. The dashed lines corresponds to $\omega_2 \hatP_0/2 \pi = 62,
  63, 64$, the dot-dashed lines to $\omega_3 \hatP_0/2 \pi = 51, 52,
  53$, and the solid lines to resonances for $\omega_1$.}
\label{fig:PhaseSpaceMultiMode}
\end{figure*}

Figure \ref{fig:PhaseSpaceMultiMode} shows two examples similar to
Fig.~\ref{fig:PhaseSpace} that explore the parameter space of
systems with three modes --- one with a dominant mode, and another with
$(\Delta a_\alpha)$ roughly equal for all modes. 
For application to stellar binaries, the example with a dominant mode
is characteristic of a binary with $M\sim (\text{a few})M_\odot$ and a
small pericentre separation $(\eta \sim 3)$. For systems with larger
$\eta$'s, the tidal potential tends to excite higher-order modes to
similar amplitudes. Appendix \ref{sec:ApE} provides more detail on the
choice of mode properties, which are determined using MESA stellar
models and the non-adiabatic GYRE pulsation code
\citep{Paxton11,Townsend13}. In general, including multiple modes does
not alter the classes of behaviours that the system exhibits. However,
the multiple-mode model is more prone to chaotic
evolution. Additionally, all modes, even those with relatively small
$\Delta a_\alpha$ can guide the evolution of the system near resonance.
	
\section{Summary and Discussion}\label{Summary}
We have developed a mathematically simple model that
	accurately captures the evolution of eccentric binary systems driven
	by dynamical tides. This model is exact for linear tidal
	oscillations in highly eccentric systems (see the last paragraph of
	Section \ref{mapping} for the regime of validity of the model). The
evolution of the ``eccentric orbit + oscillation mode'' system can be
described by an iterative
map, and depends on three parameters (for a single mode system):
$\hatP_0$, $|\Delta \hatP_1|$ and $\hatgamma$ [see
equations~(\ref{eq:defP0hat})-(\ref{eq:defgammahat})], corresponding
to the initial orbital period, the change in orbital period during the
first pericentre passage, and the damping rate of an oscillation
mode. Multiple modes can be easily incorporated.

The iterative map reveals the following key findings:
\begin{itemize}
	
	\item For non-dissipative systems, the mode evolution exhibits three
	types of behaviours, depending on the values of $|\Delta \hatP_1|$
	and $\hatP_0$ (see Figs.~\ref{fig:PhaseSpace} and \ref{fig:3behaviours}): 
	\begin{enumerate}[(i)]
		\item  For small $|\Delta \hatP_1|$
	and an orbital frequency far from resonance with the mode frequency
	(i.e. $\hatP_0/2\pi$ not close to an integer), the mode experiences
	low-amplitude oscillations with a maximum mode energy given by
	equation~(\ref{eq:EmaxOsc}). \\
	
		\item  For small $|\Delta \hatP_1|$ and
	near resonance ($\hatP_0/2\pi$ close to an integer), the mode
	exhibits larger-amplitude oscillations with a maximum energy given
	by equation~(\ref{eq:EmaxRes}). \\
	
	\item For $|\Delta \hatP_1|\gtrsim
	1$, the mode energy can grow chaotically (see Fig.~4), reaching a
	maximum of order the orbital binding energy [see
	equation~(\ref{eq:EmaxChaos})]. The chaotic mode growth can be
	approximately described by a diffusion model [equation
	(\ref{eq:AvgChaoticGrowth})], although such a model would not
	contain the energy maximum.
	\end{enumerate}
	\item 
	When mode dissipation is added, all systems, even those evolving
	chaotically, decay to a quasi-steady state (see Figs.~\ref{fig:EnergyOscDamp}-\ref{fig:EnergyTamedChaos}), with the
	mode energy and orbital decay rate given by equations~(\ref{eq:QSS})
	and (\ref{eq:eTransRateOsc}), respectively.  Continued orbital decay
	is punctuated by resonances (see Section
	\ref{ResonanceDissipation}).
\end{itemize}

These results are applicable to a variety of astrophysical systems
mentioned in the introduction. In particular, a tidally captured star
around a compact object in dense clusters \citep{Mardling01}
or a massive black hole, \citep[e.g.][]{Li13}
at the centre of galaxies may experience chaotic growth of mode amplitude
during multiple pericentre passages, accompanied by significant
orbital decay and tidal heating.  A similar evolution may occur when a
giant planet (``cold Jupiter'') is excited into a high-eccentricity
orbit by an external companion (a distant star or a nearby planet) via the
Lidov-Kozai mechanism \citep{Wu03, Fabrycky07, Nagasawa08, Petrovich15, Anderson16, Munoz16}.
We have found that a $m_{p}\sim 1~M_J$ planet pushed into an
orbit with pericentre distance $\lesssim 0.015$~AU and $e\gtrsim 0.95$
will enter the chaotic regime for the growth of f-modes. The planet
can spend an appreciable time in the high-$e$ phase of the Lidov-Kozai
cycle, allowing the mode energy to climb to a large value at which the mode becomes non-linear and suffers rapid decay.  
The consequence is 
that the planet's orbit quickly shrinks (by a factor of a few),
similar to the behaviour depicted in Fig.~\ref{fig:EnergyTamedChaos},
and the system eventually enters a quasi-steady state with slow orbital decay.
We suggest that this is a promising mechanism for forming eccentric warm Jupiters,
whose origin remains poorly understood \citep{Petrovich16, Antonini16, Huang16,Anderson17}. This mechanism also speeds up the formation of hot Jupiters through high-eccentricity migration channels.

Our study has revealed a rich variety of dynamical behaviours for highly 
eccentric binaries undergoing tidal interactions. Nevertheless, our model is still idealized.
One effect we did not include is stellar (or planetary) rotation.
The qualitative behaviours of systems that undergo low-amplitude
oscillations or chaotic evolution are unlikely to change with the
inclusion of rotation. However, tidal spin-up of the star (and tidal heating) can directly affect the
mode frequencies, giving rise to the possibility of resonance locking under some
conditions, which may extend the time frame over which the orbital energy
rapidly decreases \citep{Witte99,Fuller12a,Burkart12,Fuller17}. In addition, as noted above, our assumption of linear damping may
fail in the chaotic regime; non-linear damping could lead to even more rapid orbital 
evolution and significant structural changes in the excited star or planet.
All of these issues deserve further study.

As this paper was under review, an independent work on dynamical tides
	in eccentric giant planets was submitted by \citet{Wu17}. 
	She considers the effect of chaotic f-mode
	evolution (approximately diffusive evolution) on the orbits of gas giants undergoing high-eccentricity
	migration, assuming that the f-mode damps non-linearly when its
	amplitude becomes too large. Her conclusion that dynamical tides
	rapidly shrink the orbit, overtaking secular
	migration, agrees with our results and the discussion above.
	
\section*{Acknowledgements} This work has been supported in part by NASA
grant NNX14AP31G and NSF grant AST-1715246, and a Simons Fellowship in
theoretical physics (DL).  MV is supported by a NASA Earth and Space
Sciences Fellowship in Astrophysics.  DL thanks the hospitality of the
Institute for Advanced Study (Fall 2016) where this work started.

\bibliographystyle{mnras}	
\bibliography{References}

\appendix

\section{Physical Justification for the Iterative Map}\label{sec:ApRealModel}

\def\bxi{{\mbox{\boldmath $\xi$}}}
\def\br{{\bf r}}

We present a brief derivation of the iterative map based on the hydrodynamics of forced stellar oscillations in binaries.  We consider the
tidally-excited oscillations of the primary body (mass $M$ and radius
$R$) by the companion of mass $M'$. The gravitational potential
produced by $M'$ reads
\begin{equation}
U({\bf r},t)=-GM'\sum_{lm}{W_{lm}r^l\over D^{l+1}}\,\, e^{-im\Phi(t)}
Y_{lm}(\theta,\phi),
\end{equation}
where ${\bf r}=(r,\theta,\phi)$ is the position vector (in spherical
coordinates) relative to the centre of mass of $M$,
$D(t)$ is the binary separation and $\Phi$ is the orbital true anomaly.
The dominant quadrupole terms have $l=|m|=2$ and $m=0$, for which
$W_{2\pm 2}=(3\pi/10)^{1/2}$ and $W_{20}=(\pi/5)^{1/2}$. 
For simplicity, we neglect stellar rotation [see \citet{Schenk02,Lai06}].
To linear order, the response of star $M$ to the tidal forcing frequency is specified by the Lagrangian displacement $\bxi(\br,t)$.
A free oscillation mode of frequency $\omega_\alpha$ has the form 
$\bxi_\alpha(\br,t)=\bxi_\alpha(\br)\,e^{-i\omega_\alpha t}\propto
e^{im\phi-i\omega_\alpha t}$.
We carry out phase-space expansion of $\bxi(\br,t)$ in terms of the eigenmodes \cite{Schenk02}:
\begin{equation}
\left[\begin{array}{c}
\bxi\\
{\partial\bxi/\partial t}
\end{array}\right]
=\sum_\alpha c_\alpha(t)
\left[\begin{array}{c}
\bxi_\alpha(\br)\\
-i\omega_\alpha\bxi_\alpha(\br)
\end{array}\right].
\end{equation}
The linear fluid dynamics equations for the forced stellar oscillations then reduce to the evolution
equation for the mode amplitude $c_\alpha(t)$ \citep{Lai06}:
\begin{equation}
\dot{c}_\alpha + (i\omega_\alpha + \gamma_\alpha)c_\alpha = \frac{i G M'W_{lm}Q_\alpha}{2\omega_\alpha D^{l+1}}
\, \text{e}^{-im\Phi(t)}, \label{eq:cdoteq}
\end{equation}
where $\gamma_\alpha$ is the mode (amplitude) damping rate, and
\begin{equation}
Q_\alpha=\frac{1}{M^{1/2} R^{(l-1)}}\int d^3x\,\rho \bxi_\alpha^\star\cdot\nabla (r^lY_{lm})
\end{equation} 
is the dimensionless tidal overlap integral. The eigenmode is normalized according to $\int d^3x \rho(\br) |\bxi_\alpha(\br)|^2=1$, which implies that $\xi$ has units of $M^{-1/2}$.

The general solution to equation~(\ref{eq:cdoteq}) is
\begin{equation}
c_\alpha = \text{e}^{-(i\omega_\alpha + \gamma_\alpha) t} \int_{t_0}^t
\frac{iGM'W_{lm}Q_\alpha}{2\omega_\alpha  D^{l+1}}\,\text{e}^{(i\omega_\alpha  +
\gamma_\alpha) t'-im\Phi(t')}\,dt'.
\end{equation}
For eccentric binaries, we can write this as a sum over multiple
pericentre passages. To do so, let $t_k$ be the time of the $k$-th apocentre
passage. [Note that, for the moment, this use of $k$ differs from the
meaning used in equations~(\ref{eq:a-})-(\ref{eq:ak}),
where it is used to describe the number of pericentre passages]. We
can relate $t_k$ to $t_{k-1}$ via
\begin{equation}
t_k = t_{k-1} + \frac{1}{2}(P_{k-1}+P_k).
\end{equation}
Now define
\begin{equation}
\Delta c_\alpha = \int_{-P_{k-1}/2}^{P_{k}/2} \frac{iGM'W_{lm}Q_\alpha}{2\omega_\alpha D^{l+1}}
\,\text{e}^{(i\omega_\alpha  + \gamma_\alpha) t'-im\Phi(t')}\;dt'.\label{eq:Deltacdef}
\end{equation}
This is the change in mode amplitude during a pericentre passage, and
it is approximately constant for any $k$ provided the orbit is 
very eccentric and the pericentre distance $r_{\rm{peri}}$ remains constant (see the main text for
discussion). We can then write $c_\alpha$ at time $t_k$ simply as
\begin{equation}
c_{\alpha,k} = \text{e}^{-(i\omega_\alpha + \gamma_\alpha) t_k}\Delta c_\alpha
\sum_{j=1}^k\,e^{(i\omega_\alpha + \gamma_\alpha)
(t_{j-1} + P_{j-1}/2)}. \label{eq:cdiscrete}
\end{equation}
We can reorganize equation~(\ref{eq:cdiscrete}) into an iterative form
\begin{equation}
c_{\alpha,k} = c_{\alpha,k-1}\,e^{-(i\omega_\alpha + \gamma_\alpha)(P_{k-1} + P_k)/2}+ 
e^{-(i\omega_\alpha + \gamma_\alpha)P_k/2}\Delta c_\alpha.\label{eq:apocenterMap}
\end{equation}
We now shift the index $k$ to count pericentre passages rather than apocentre passages by defining
\begin{equation}
a_{\alpha,k} =\sqrt{2}\omega_\alpha c_{\alpha,k}\,\text{e}^{-i\omega P_k/2}/|E_{B,0}|,
\end{equation} 
where we have also re-normalized the mode amplitude so that the scaled mode energy (in units of $|E_{B,0}|$) is $\tildeE_{\alpha,k} = |a_{\alpha,k}|^2.$ (Note the mode energy is given by $E_{\alpha,k} = 2\omega_\alpha^2 |c_{\alpha,k}|^2$.) Equation~(\ref{eq:apocenterMap}) then reduces to 
\begin{equation}
a_{\alpha,k} = (a_{\alpha,k-1}+ \Delta a_\alpha)\,e^{-(i\omega_\alpha + \gamma_\alpha)P_k},
\end{equation}
where $\Delta a_\alpha = \sqrt{2}\omega_\alpha \Delta c_\alpha/|E_{B,0}|$.

Using equation~(\ref{eq:Deltacdef}), we can write $\Delta E_{\alpha,1}$ explicitly in terms of orbital parameters and mode properties:
\begin{equation}
\Delta E_{\alpha,1} = |E_{B,0}|(\Delta a_\alpha)^2 = \frac{GM'^2}{R}\left(\frac{R}{r_{\rm peri}}\right)^{2(l+1)} T(\eta,\omega_\alpha/\Omega_{\rm peri},e), \label{eq:DeltaE1}
\end{equation}
where the dimensionless function $T$ is given by
\begin{equation}
T = 2\pi^2 Q_\alpha^2 K_{lm}^2. \label{eq:Tdef}
\end{equation}
Ignoring the negligible effect of mode damping at pericentre, we have 
\begin{equation}
K_{lm} = \frac{W_{lm}}{2 \pi} \left(\frac{GM}{R^3}\right)^{1/2} \int_{-P/2}^{P/2} dt' \left(\frac{r_{\rm peri}}{D}\right)^{l+1} \text{e}^{\text{i}\omega_\alpha t' - i m \Phi(t')}. \label{eq:Kdef}
\end{equation}
The energy transfer for a parabolic passage $(e \to 1)$ was first derived in \citet{Press77}. Equations~(\ref{eq:DeltaE1})-(\ref{eq:Kdef}), which apply to eccentric orbits as well, were presented in \citet{Lai97} [see equations (22)-(23)] and in \citet{Fuller12a} [see
equation (14)-(15); note that in equation~(15), $R$ should be replaced by $D_p$.]

Note that the integral in equation~(\ref{eq:Kdef}) is difficult to calculate accurately and efficiently because the mode frequency is typically orders of magnitude larger than the orbital frequency. However, for parabolic orbits, $K_{lm}$ can be evaluated in the limit $\omega_\alpha/\Omega_{\rm peri} \gg 1$ \citep{Lai97}. For example, for $l=2$, $m=2$, 
\begin{equation} 
K_{22} = \frac{2 z^{3/2}\eta^{3/2}\text{e}^{-2z/3}}{\sqrt{15}}\left(1 - \frac{\sqrt{\pi}}{4\sqrt{z}}+\cdots\right), \label{eq:Kapprox}
\end{equation}
	where
\begin{equation}
	z\equiv \sqrt{2} \omega_\alpha/\Omega_{\rm peri}.
\end{equation}
This expansion approximates $K_{lm}$ to within a few percent for $(1-e)<< 1$ and $z \gtrsim 3$. For typical f-mode frequencies, the latter condition is satisfied for $\eta \gtrsim$~a few.

\section{Non-dissipative Systems}\label{sec:ApA}

\subsection{Maximum Mode Energy for Non-Chaotic Systems}\label{sec:ApRes}

In the oscillatory and resonant regimes, the mode energy $\tilde E_k\ll 1$, and the iterative
map given by equation~(\ref{eq:maplinear}) can be rewritten:
\begin{equation}
z_{k+1}\simeq 1+z_k\, e^{-i\hatP_0+i|\Delta\hatP_1| |z_k|^2},
\end{equation}
where $z_k\equiv a_{k-}/\Delta a=a_{k-1}/\Delta a +1$.

{\bf Oscillatory Regime}: When $|\delta \hatP_0|=|\hatP_0-2\pi N|\gg |\Delta\hatP_1| |z_k|^2$, the map 
simplifies to 
\begin{equation}
z_{k+1}\simeq 1+z_k\, e^{-i\hatP_0}.
\end{equation}
This yields the solution (for $z_1=1$)
\begin{equation}
z_k\simeq {1-e^{-ik\hatP_0}\over 1-e^{-i\hatP_0}},
\end{equation}
which is equivalent to equation~(\ref{eq:OscSolution}).  
The validity of this oscillatory solution requires $|\delta\hatP_0|\gg |\Delta\hatP_1|/(1-\cos\hatP_0)$ or
$|\delta\hatP_0|^3\gg |\Delta\hatP_1|$.

{\bf Resonant Regime}: When $|\delta\hatP_0|^3\ll |\Delta\hatP_1|$, the system is in the resonant
regime, and the map becomes
\begin{equation}
z_{k+1}-z_k \simeq 1 + i|\Delta\hatP_1| |z_k|^2 z_k.
\end{equation}
For $k\gg 1$, we can approximate the mode amplitude as a continuous function of $k$, and the
above equation reduces to 
\begin{equation}
\frac{dz}{dk} \simeq 1 + i |\Delta \hatP_1| |z|^2 z. \label{eq:dydk}
\end{equation}
Now we express $z$ explicitly in terms of an amplitude $A$ and phase $\theta$:
\begin{equation}
z  = A\,\text{e}^{\text{i} \theta}.
\end{equation}
Equation~(\ref{eq:dydk}) can be rewritten as two differential equations:
\begin{align}
\frac{dA}{dk} &\simeq \cos\theta, \label{eq:dthetadk}\\
\frac{d\theta}{dk} &\simeq  \frac{1}{A}\left(|\Delta \hatP_1| A^3 - \sin\theta \right).
\end{align}
We combine these to examine how the amplitude varies with the phase:
\begin{align}
\frac{dA}{d\theta} = \frac{A \cos\theta}{\left(|\Delta \hatP_1| A^3 - \sin\theta \right)}. \label{eq:dAdtheta}
\end{align}
To solve the above equation, we use the substitutions 
\begin{equation}
u=|\Delta \hatP_1| A^3, \quad v=\sin\theta.
\end{equation}
Equation~(\ref{eq:dAdtheta}) then simplifies to
\begin{equation}
\frac{du}{dv} = \frac{3u}{u-v}.
\end{equation}
For the initial condition $u=v=0$, which corresponds to $a_0 = 0$, the solution is simply $u = 4v$, or
\begin{equation}
A = \left(\frac{4 \sin \theta}{|\Delta \hatP_1|}\right)^{1/3} \label{eq:Aoftheta},
\end{equation}
which has the maximum value $(4/|\Delta \hatP_1|)^{1/3}$. 
The maximum mode energy for a system near resonance is therefore
\begin{equation}
{\tilde E}_{\rm res}\equiv \tilde{E}_{k,\text{max}}=
(\Delta a)^2\left(\frac{4}{|\Delta \hatP_1|}\right)^{2/3} \simeq
\frac{2^{7/3}}{3}{|\Delta\hatP_1|^{1/3}\over\hatP_0}.
\end{equation}

We can use the above result to approximate the shape that the mode amplitude traces in the complex 
plane over many orbits. Our numerical calculation (see Fig. \ref{fig:3behaviours}) shows that the mode amplitude 
as a function of its phase can be described by 
\begin{equation}
\left|\frac{a_k}{\Delta a} + 0.5\right| \simeq 
\left(\frac{4 \sin \theta}{|\Delta \hatP_1|}\right)^{1/3} ,\label{eq:akRes}
\end{equation}
where $\theta$ is the phase of $a_k/\Delta a +0.5$. This is similar to equation~(\ref{eq:Aoftheta})
except for a shift along the real axis.
For $|\Delta \hatP_1|\ll 1$, this approximation performs very well, as seen in
Fig.~\ref{fig:3behaviours}.

\subsection{Resonant Timescale}\label{sec:ApResTimescale}

The mode of a non-dissipative system near resonance evolves 
periodically, repeatedly tracing out a closed shape in the complex
amplitude plane. We define $t_{\text{res}}$ as the period of the resonant
oscillations. To calculate this timescale, we use equations~(\ref{eq:dthetadk}) and (\ref{eq:Aoftheta}) to find 
\begin{equation}
\frac{d\theta}{dk} \simeq \left(\frac{|\Delta \hatP_1|}{4}\right)^{\!1/3} 3 (\sin\theta)^{2/3}.
\end{equation}
Integrating the above differential equation from $\theta=0$ to $\theta=\pi$ gives
\begin{equation}
t_{\text{res}} \simeq 3.85 \; P |\Delta \hatP_1|^{-1/3}, 
\end{equation}
where $P$ is the orbital period associated with the resonance.
In order of magnitude, the number of orbits necessary to reach the maximum mode amplitude $|a_{\text{res}}| \sim |\Delta \hatP_1|^{-1/3} \Delta a$ is simply $|a_{\rm{res}}|/ \Delta a$.

\subsection{Maximum Mode Energy for Chaotic Systems} \label{sec:ApC}

As discussed in the main text, the mode energy of a chaotic system initially grows stochastically with an expected value of $\langle \tildeE_k \rangle \sim \Delta \tildeE_1 k$ [see equation~(\ref{eq:AvgChaoticGrowth}) and Fig. \ref{fig:RandomPhase}], but cannot exceed a maximum value, $\tildeE_{\rm{max}}$. As the mode energy increases, the change in the orbital period between pericentre passages, $\Delta \hat{P}_k$, decreases [see equation~(\ref{eq:Pdiff})]. The maximum mode energy is approximately set by $|\Delta \hatP_k| \sim 1$, and is given by equation~(\ref{eq:EmaxChaos}). 

This condition is related to that found in \citet{Mardling95a,Mardling95b}, where the maximum mode energy is set by a ``chaos boundary" that separates orbital parameters that produce chaotic behaviour from those that produce oscillatory behaviour. The location of the boundary depends on the current mode amplitude. The iterative map in this paper demonstrates that such boundaries are determined by the size of $|\Delta \hatP_k|$, which depends on $\Delta a$ and $a_{k-1}$ [see equation~(\ref{eq:Pdiff})]. Both the onset of chaotic behaviour and the maximum mode energy are set by conditions on $|\Delta \hatP_k|$ (where $k=1$ when considering the onset). It follows that both conditions are related to ``chaos boundaries", as observed by \citet{Mardling95a,Mardling95b}.

\begin{figure}
	\begin{center}
	\includegraphics[width=\columnwidth]{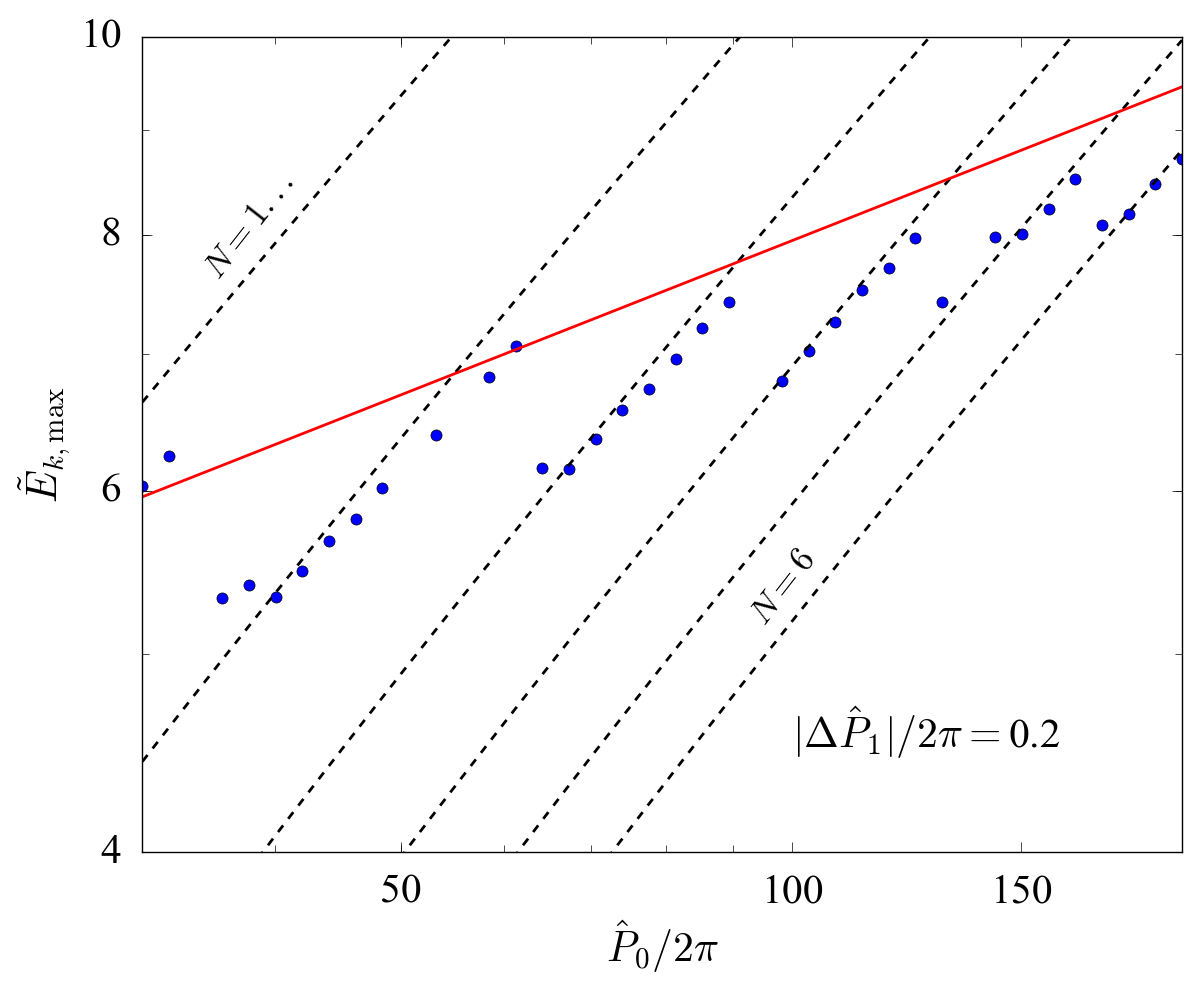}
	\caption{Numerical results for $\tilde{E}_{\rm{max}}$ (blue dots) after $3 \times 10^6$ orbits for chaotic systems with $|\Delta\tilde{P}_1|/2 \pi = 0.2$. The red line is $1.5 (\hatP_0|\Delta \hatP_1|)^{1/4}$ [see equation~(\ref{eq:EmaxChaos})]. The dotted lines are from equation~(\ref{eq:EmaxvsN}) with different values of $N$.}
	\label{fig:EmaxvsP0Chaos}
	\end{center}
\end{figure}

\begin{figure}
	\begin{center}
	\includegraphics[width=\columnwidth]{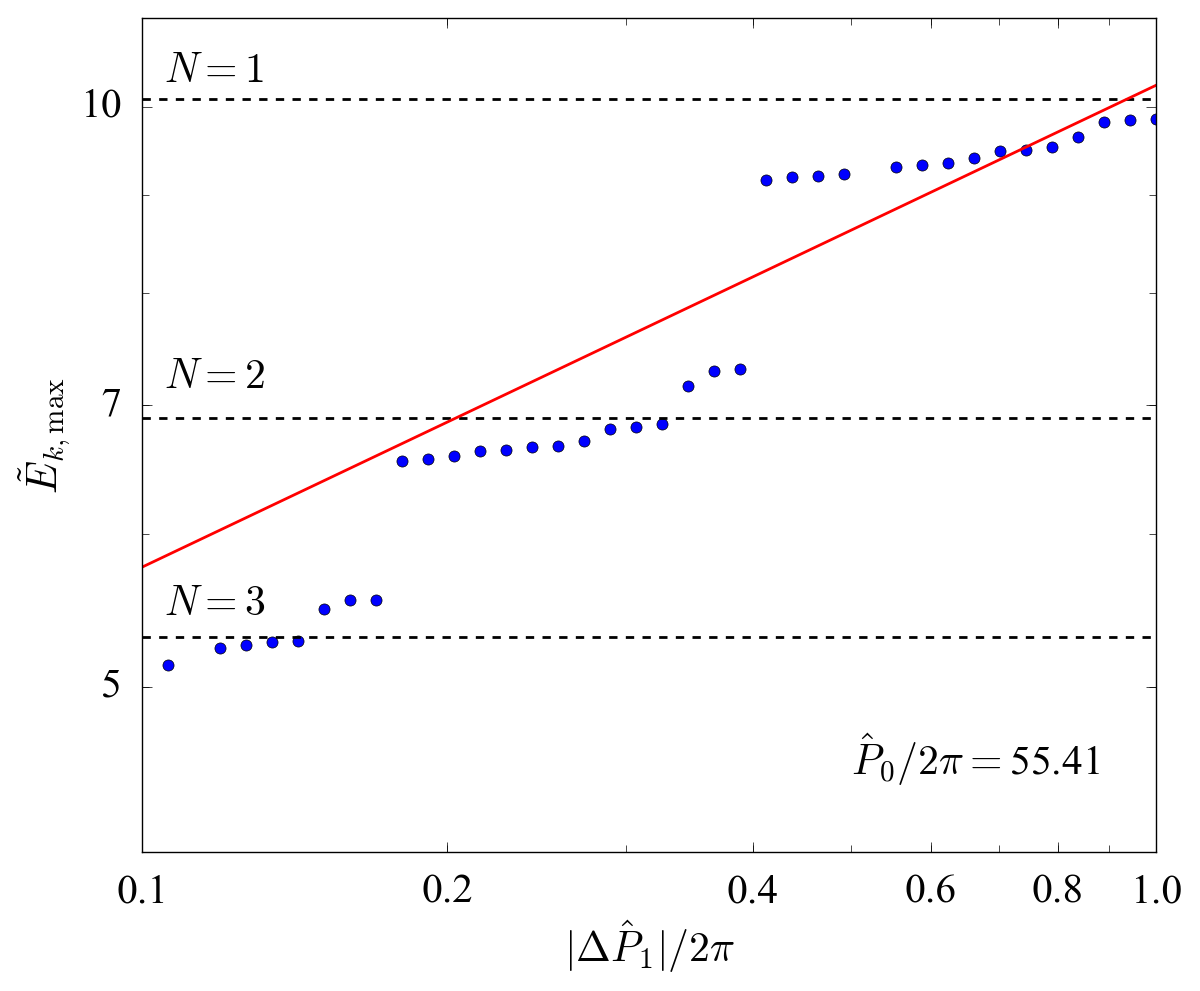}
	\caption{Numerical results for $\tilde{E}_{\rm{max}}$ (blue dots) after $3 \times 10^6$ orbits for chaotic systems with $\hatP_0/2 \pi = 55.41$. The red line is $1.5 (\hatP_0|\Delta \hatP_1|)^{1/4}$ [see equation~(\ref{eq:EmaxChaos})]. The dotted lines are from equation~(\ref{eq:EmaxvsN} with different values of $N$).}
	\label{fig:EmaxvsDelPChaosApprox}
	\end{center}
\end{figure}

In reality, the dependence of $\tildeE_{\rm{max}}$ on $\hatP_0$ and $|\Delta \hatP_1|$ is more complicated than the power law trend from equation~(\ref{eq:EmaxChaos}). Figures \ref{fig:EmaxvsP0Chaos} and \ref{fig:EmaxvsDelPChaosApprox} show ``jumps" and ``drops" in $\tildeE_{\rm{max}}$ at some values of $\hatP_0$ and $|\Delta \hatP_1|$. To understand this step-like behaviour, we note that the maximum mode energy for a non-dissipative system is associated with the minimum (dimensionless) orbital period by
\begin{equation}
	\hatP_{\rm{min}} = \hatP_0\left(\frac{1}{1+\tildeE_{\rm{max}}}\right)^{3/2}.
\end{equation}
We have found that for a chaotic system, as $|\Delta \hatP_k|$ decreases toward unity, the orbital period tends to evolve away from resonances with the stellar mode and to linger directly between them. This behaviour can be understood from the fact that, for a system near resonance, the shifts in mode amplitude during pericentre tend to add over successive passages, pushing the system away from the resonance. Imposing $\hatP_{\rm{min}} \simeq 2\pi(N + 1/2)$ yields
\begin{equation}
	\tildeE_{\rm{max}}\simeq \left[\frac{\hatP_0}{2\pi(N+1/2)}\right]^{2/3} -1.
	\label{eq:EmaxvsN}
\end{equation}
Equation~(\ref{eq:EmaxvsN}) is in good agreement with results shown in Figs. \ref{fig:EmaxvsP0Chaos} and \ref{fig:EmaxvsDelPChaosApprox}. We see that the jumps in $\tildeE_{\rm{max}}$ correspond to changes in $N$. Combining equation~(\ref{eq:EmaxvsN}) with the broader trend of equation~(\ref{eq:EmaxChaos}) captures the main features of how $\tildeE_{\rm{max}}$ depends on $\hatP_0$ and $|\Delta \hatP_1|$ for chaotic systems. 

\section{G-mode Properties of Stellar Models}\label{sec:ApE}

\begin{figure}
\begin{center}

\includegraphics[width=\columnwidth]{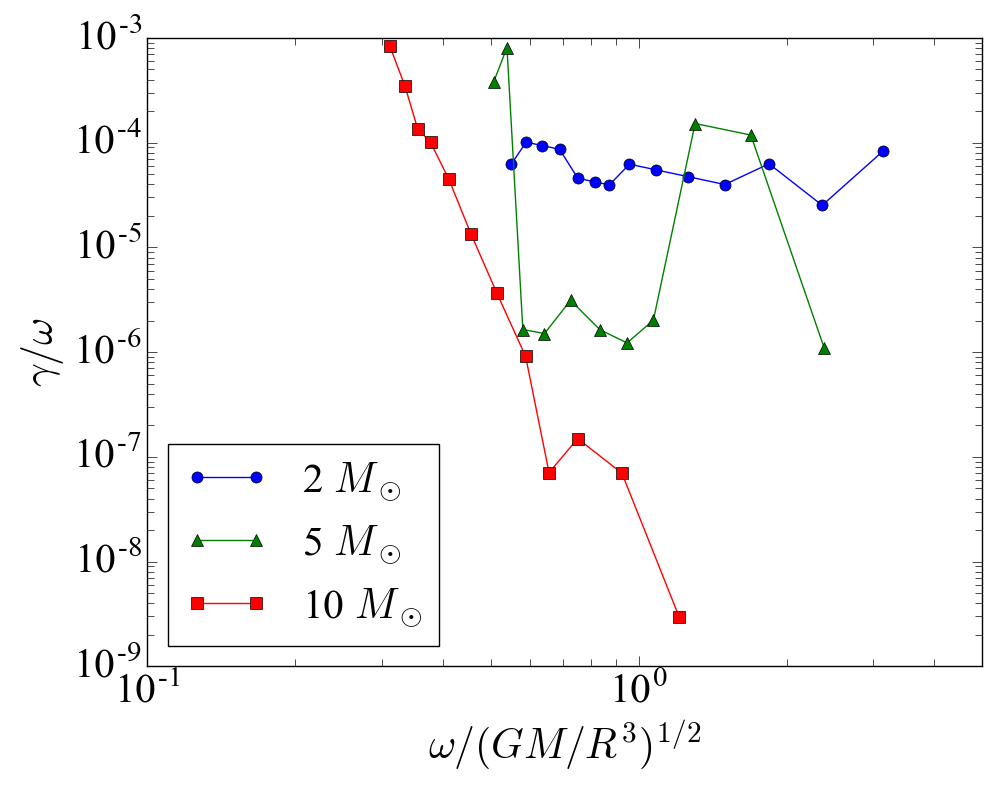}
\caption{Numerical results for the g-mode frequencies ($\omega$) and
  damping rates ($\gamma$) of three MESA stellar models analysed with the
  non-adiabatic stellar pulsation code, GYRE.  The dip in
  $\gamma/\omega$ for the $5 M_\odot$ model is typical for models in
  the mass range $4-8 M_\odot$.}
\label{fig:gammaHatMultipleModels}
\end{center}
\end{figure}

One application of our model is the tidal
capture of main-sequence stars by compact objects, including massive black holes. We use the stellar evolution
code, MESA \citep{Paxton11}, and
the non-adiabatic pulsation code, GYRE \citep{Townsend13}, to determine
the properties of g-modes in the radiative envelope of stars between
$2$ and $10 M_\odot$. The characteristic damping times of modes are
found using the imaginative part of the mode frequency. These values
are generally in good agreement with a quasi-adiabatic approximation
that assumes radiative damping.
Figure \ref{fig:gammaHatMultipleModels} shows the
computed damping rates for three stellar models. For the $2 M_\odot$
model, the damping rate only varies by a factor of a few for the relevant g-modes. For
the $10 M_\odot$ model (and other models with $10 M_\odot \le M \le 20
M_\odot$), the damping rates are much smaller for higher frequency modes.

\begin{figure}
	\begin{center}
	\includegraphics[width=\columnwidth]{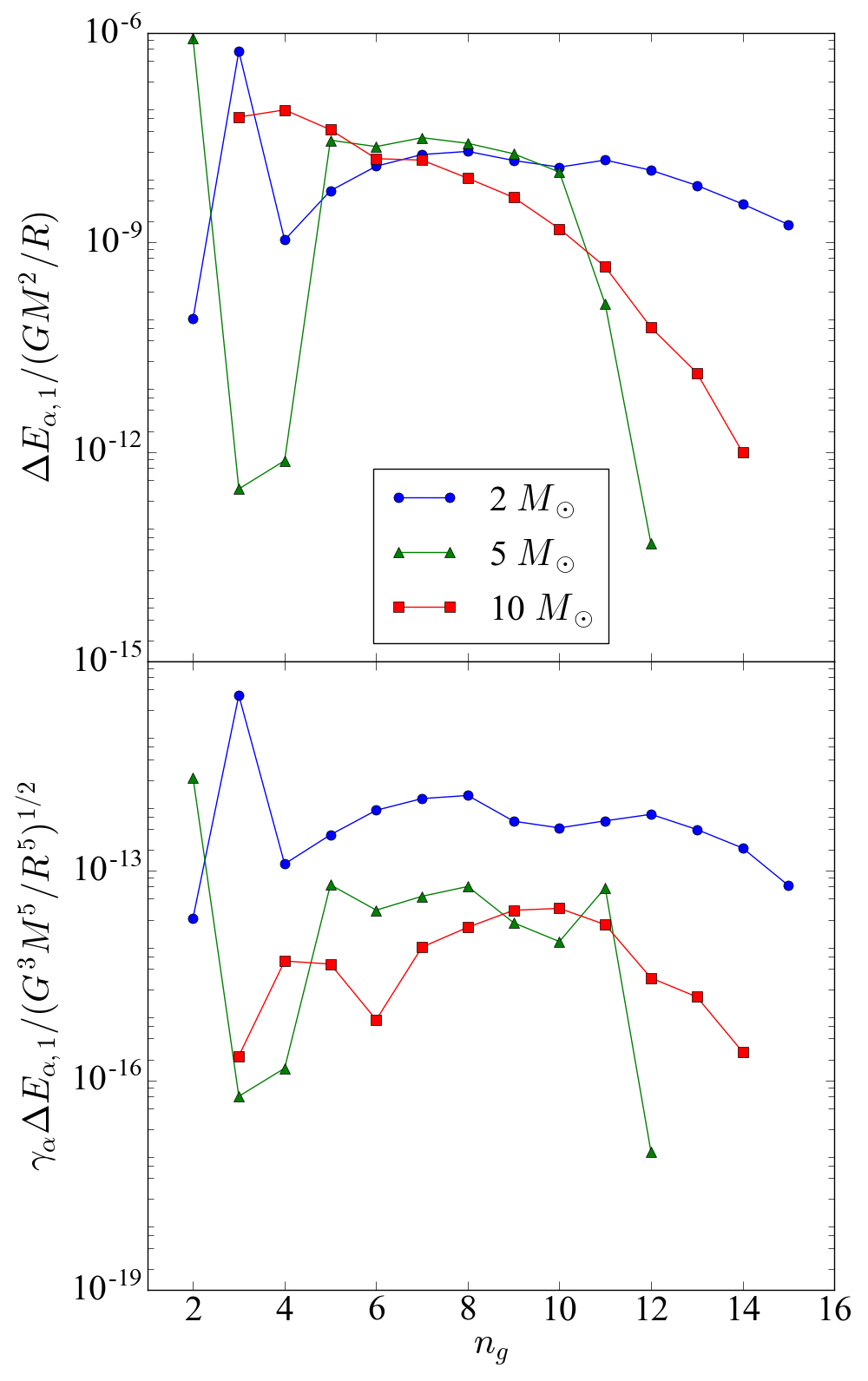}
	\caption{Numerical results for $\Delta E_{\alpha,1}$ and $\gamma_\alpha \Delta E_{\alpha,1}$ vs. $n_g$ (the radial mode number) for three MESA stellar models analysed with the non-adiabatic stellar pulsation code, GYRE. The energy transfer $\Delta E_{\alpha,1}$ is calculated assuming $\eta = 3$ and $e=0.95$.}
	\label{fig:DeltaEMultipleModels}
	\end{center}
\end{figure}

The amount of energy transferred to a mode (labelled $\alpha$) in the ``first'' pericentre passage 
$\Delta E_{\alpha,1}$ (i.e., when the oscillation amplitude is zero before the passage)
depends on the stellar structure and mode properties.
We use the method of \citet{Press77}
to calculate $\Delta E_{\alpha,1}$. The quasi-steady-state mode dissipation rate from
equation~(\ref{eq:QSS}) is of order $\gamma_\alpha\Delta E_{\alpha,1}$. Figure \ref{fig:DeltaEMultipleModels} shows
an example of the calculated $\Delta E_{\alpha,1}$ and the energy dissipation 
rates for systems with different stellar properties and $\eta = 3$.
We find that stars with $M \lesssim 5 M_\odot$ tend to have a single low-order g-mode with large $\Delta \tildeE_{\alpha,1}$ that dominates the energy transfer rate. To represent a system with a dominant mode, we choose the $\omega$ and $\gamma$ ratios between modes from the $2 M_\odot$ model and the  $\tildeE_{\alpha,1}$ ratio for $\eta = 3$. More massive stars ($M \gtrsim M_\odot$) have a number of g-modes that contribute roughly equally to the energy transfer rate. To represent a system where multiple modes are important for energy transfer, we choose the $\omega$ and $\gamma$ ratios between modes from the $10 M_\odot$ model and the  $\tildeE_{\alpha,1}$ ratio for $\eta = 3$.

\begin{figure}
	\begin{center}
	\includegraphics[width=\columnwidth]{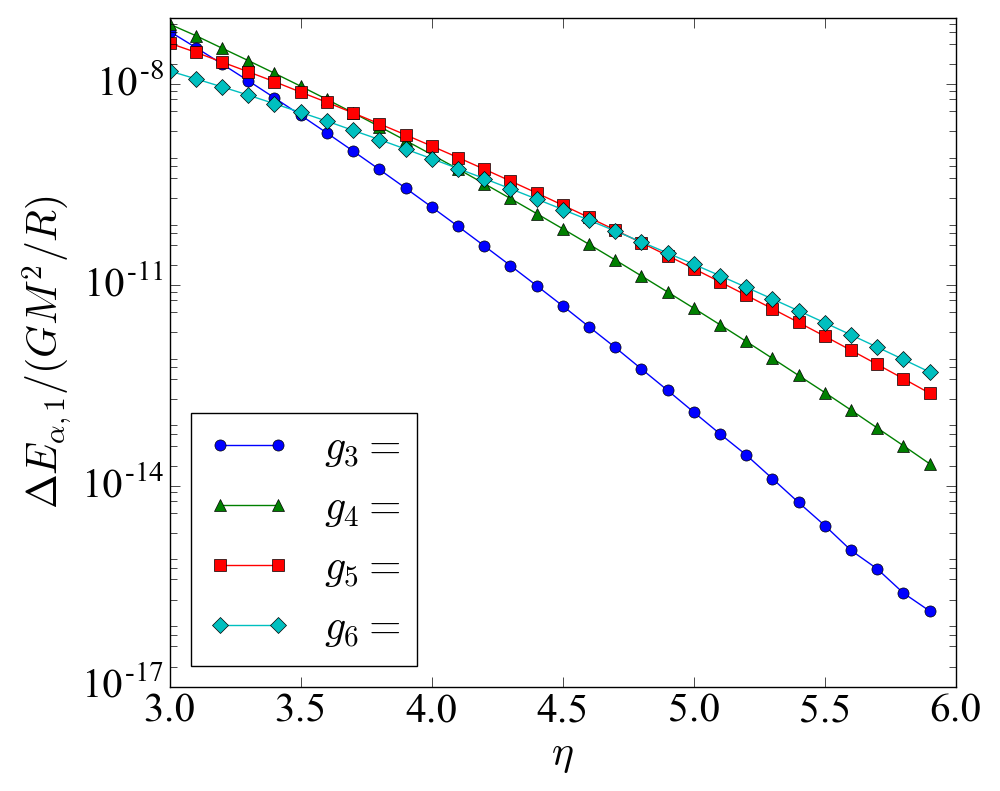}
	\caption{Numerical results for $\Delta E_{\alpha,1}$ for four g-modes of the $10 M_\odot$ stellar model as a function of $\eta$. For larger $\eta$, higher order g-modes receive more energy at pericentre than lower-order g-modes. }
	\label{fig:ModeVaryEta}
	\end{center}
\end{figure}

The orbital parameter $\eta=(r_{\rm{peri}}/R)(M/M_t)^{1/3}$ (where 
$r_{\rm{peri}}$ is the pericentre distance) also strongly affects $\Delta E_{\alpha,1}$, though the dependence on $e$ is negligible for highly eccentric
orbits. 
For larger $\eta$, the orbital frequency at pericentre is smaller, and higher-order g-modes
contribute more to tidal energy transfer, as illustrated in Fig. \ref{fig:ModeVaryEta}.

\bsp
\label{lastpage}
\end{document}